\documentclass[prd,twocolumn,floatfix,superscriptaddress,nofootinbib]{revtex4-1}
\usepackage{dcolumn,amsmath}
\usepackage{graphicx}
\usepackage{braket}
\usepackage{bm}
\usepackage{amssymb}
\usepackage{hyperref}
\usepackage{multirow}
\usepackage{footnote}
\usepackage{subcaption}
\usepackage{ragged2e}
\usepackage{xcolor}
\usepackage{pdfcomment}
\usepackage[justification=justified,labelfont=bf,singlelinecheck=off]{caption}
\usepackage{threeparttable,booktabs}  
\usepackage{subcaption,siunitx,booktabs}
\usepackage{caption,afterpage,tabularx}
\usepackage{diagbox}
\usepackage{mathrsfs}
\usepackage{physics}
\usepackage{cancel}
\usepackage{amsmath}
\newcommand{\RomanNumeralCaps}[1]
{\MakeUppercase{\romannumeral #1}}

\usepackage[title]{appendix}

\newcommand{\JSU}{
School of Physics and Electronic Engineering,
\\Jiangsu University, 301 Xuefu Rd., Zhenjiang, Jiangsu, China\\
}

\newcommand{\MUST}{State Key Laboratory of Lunar and Planetary Sciences, 
\\Macau University of Science and Technology, Macao, China}

\begin{document}

\title{Neutron-neutral particle mixing and its observable consequences}
\author{Yongliang Hao}
%\email{yhao@ujs.edu.cn}
\affiliation{\JSU}
\affiliation{\MUST}
\author{Dongdong Ni}
\email{ddni@must.edu.mo}
\affiliation{\MUST}

%\affiliation{\RUG}

\date{\today}

\begin{abstract}

In this work, we explore the mixing between neutron ($n$) and elementary neutral particle ($\eta$), which violates both the baryon number ($\mathcal{B}$) and the lepton number ($\mathcal{L}$) by one unit but conserves their difference $(\mathcal{B}-\mathcal{L})$. Such mixing may give rise to non-trivial effects that are different from the Standard Model predictions. We organize our discussions based on two scenarios, roughly depending on whether an interference between oscillation and decay occurs, or whether the new-physics effects associated with the $n$-$\eta$ mixing contribute to the absorptive mixing amplitude. If an oscillation process is not accompanied by an interference between oscillation and decay, or the new-physics interactions do not contribute to the absorptive mixing amplitude, such a process can be classified as pure oscillation (e.g. neutrino oscillation). Otherwise, it can be classified as impure oscillation (e.g. meson-antimeson oscillation). In the scenario of pure oscillation, CP-violation arising from the Majorana phase can manifest itself through the $n$-$\bar{n}$ oscillation process and may lead to observable effects. In the scenario of impure oscillation, we analyze the testable implications on the masses and lifetimes of the mass eigenstates formed as a result of the $n$-$\bar{n}$ oscillation mediated by $\eta$. In this scenario, we also suggest a unified interpretation of the neutron lifetime anomaly and the $n$-$\bar{n}$ oscillation measurements based on the $n$-$\eta$ mixing. In both scenarios, we present the lower bounds imposed by the experimental searches for $n$-$\bar{n}$ oscillations on the masses of the color multiplet bosons and point out that they could be within the reach of a direct detection at the LHC or future high-energy experiments. Furthermore, we discuss about the observability of the geometric phase associated with the $n$-$\eta$ mixing. The measurement of such a geometric phase may provide another opportunity for the study of new-physics effects.

\end{abstract}

\maketitle

\section{Introduction\label{sec1}}

New physical phenomena beyond the Standard Model (SM) have been mathematically predicted by many new physics models and intensively explored in a wide variety of experiments over the past decades \cite{zyla2020review}. Since many new physics models are featured with particle mixing and oscillation, the phenomena of particle mixing and oscillation play a critical role in the construction of the extensions to the SM. For example, the neutrino flavor oscillation has been confirmed in various scenarios \cite{fukuda1998evidence,fukuda2000tau,ahmad2001measurement,agafonova2010observation,an2012observation,abe2013evidence} and indicates that at least one type of neutrino has a non-zero mass, which contradicts the basic assumption of the SM and suggests that the SM is not perfect and thus new physics models need to be constructed \cite{zyla2020review}. Furthermore, particle mixing and oscillation are indispensable to understand CP-violating effects, which have been confirmed by many experiments (see e.g. Refs. \cite{christenson1964evidence,aubert2001observation,abe2001observation,aaij2013first,chala2019deltaa}).

Cold neutrons can serve as a rich and varied environment where many interesting processes occur \cite{snow2022searches}, making it possible to search for new physical phenomena in a smaller experiment, comparing with the ones at the LHC. The neutron lifetime and the neutron-antineutron ($n$-$\bar{n}$) oscillation time are the two key observables that are investigated intensively \cite{baldo1994new,jones1984search,takita1986search,berger1990search,chung2002search,aharmim2017search,abe2015search,abe2021neutron,czarnecki2018neutron,gonzalez2021improved}. Theoretical investigations on the properties of neutron not only can help develop an efficient measurement strategy for new physical phenomena, but also can help interpret the results more correctly after the measurements. In this work, we focus on the theoretical aspects associated with the measurements of the neutron lifetime and the $n$-$\bar{n}$ oscillation.

The $n$-$\bar{n}$ oscillation, which violates baryon number ($\mathcal{B}$) by two units, has attracted an enormous level of attention both theoretically and experimentally \cite{phillips2016neutron}. The searches for the $n$-$\bar{n}$ oscillation have been performed in various mediums \cite{phillips2016neutron}, including bound states, field-free vacuum, and etc. Up to date, no significant signal for the $n$-$\bar{n}$ oscillation has been found. In field-free vacuum, the lower limit on the $n$-$\bar{n}$ oscillation time presented by the Institut Laue-Langevin (ILL) experiment is approximately $0.86 \times 10^{8}$ s \cite{baldo1994new}. In bound states, the $n$-$\bar{n}$ oscillations have been searched for by various experiments, such as Irvine-Michigan-Brookhaven (IMB) \cite{jones1984search}, Kamiokande (KM) \cite{takita1986search}, Frejus \cite{berger1990search}, Soudan-2 (SD-2) \cite{chung2002search}, Sudbury Neutrino Observatory (SNO) \cite{aharmim2017search}, Super-Kamiokande (Super-K) \cite{abe2015search,abe2021neutron}, and etc. Among them, the most stringent constraint on the $n$-$\bar{n}$ oscillation time is imposed by the Super-K experiment with the value of $4.7 \times 10^{8}$ s \cite{abe2021neutron}, when converting to the field-free vacuum values. Although the measurements on neutrons bound in nuclei provide relatively tighter limits, such limits depend heavily on the details of the nuclear models \cite{baldo1994new}, whereas the limits imposed by the measurements on free neutrons are model independent.

From the theoretical aspect, the $n$-$\bar{n}$ oscillation can be predicted by many new physics models (see e.g. Ref. \cite{mohapatra2009neutron}), such as left-right symmetry model \cite{mohapatra1980local,babu2009neutrino}, grand unified symmetry model \cite{lust1982neutron,kalara1983supersymmetric}, super-symmetry model \cite{kalara1983supersymmetric,chacko1999supersymmetric,dutta2006observable,babu2007unified}, extra dimension model \cite{nussinov2002n,mohapatra2009neutron}, mirror world model \cite{berezhiani2012magnetic,berezhiani2021possible,berezhiani2021neutron}, and etc. Among such models, the mirror world model was initially proposed for the understanding of parity violation \cite{lee1956question,kobzarev1966on,foot1991model}. As a special case of the mirror world model, the mixing between neutron ($n$) and mirror neutron ($n^{\prime}$) has been studied \cite{berezhiani2004mirror,mohapatra2005some,berezhiani2006neutron,foot2014mirror,berezhiani2019neutron}. Such a mixing may give rise to new physical phenomena (see e.g. Ref. \cite{addazi2021new}), such as neutron disappearance ($n$-$n^{\prime}$) \cite{berezhiani2019neutron2,berezhiani2021neutron,babu2022theoretical}, neutron regeneration  ($n$-$n^{\prime}$($\bar{n}^\prime$)-$n$) \cite{berezhiani2009more,berezhiani2017neutron,kamyshkov2022neutron}, and neutron-antineutron oscillation [$n$-$n^{\prime}$($\bar{n}^\prime$)-$\bar{n}$] \cite{mohapatra2005some,berezhiani2006neutron,berezhiani2016neutron,berezhiani2021possible}. Instead of direct mass mixing terms, the $n$-$\bar{n}$ oscillation can be achieved indirectly through mirror particles as intermediate states \cite{berezhiani2016neutron,berezhiani2021possible}. Recently, high-sensitivity measurement schemes for the $n$-$n^{\prime}$ oscillation has been designed and demonstrated \cite{addazi2021new,ayres2022improved}.

The neutron lifetime anomaly, which refers to the discrepancy in the measured neutron lifetime between two different experimental approaches, has attracted great attention recently (see e.g Ref. \cite{zyla2020review}). For example, the trap experiment reports a neutron lifetime $\tau_n = 877.75^{+0.28}_{-0.28} \text{(stat.)}^{+0.22}_{-0.16}\text{(syst.)}$ s \cite{gonzalez2021improved} through the observation of neutron disappearance. However, the beam experiment reports a neutron lifetime $\tau_n = 887.7^{+1.2}_{-1.2} \text{(stat.)}^{+1.9}_{-1.9}\text{(syst.)}$ s \cite{yue2013improved} through the neutron $\beta$-decay products, such as protons and electrons. The results presented by the two different approaches provide an approximately 4 $\sigma$ deviation \cite{czarnecki2018neutron}, which may imply a signal for new physics. It has been shown that the interpretation of neutron lifetime anomaly based on exotic dark decay channels can be excluded through the analysis of the neutron decay $\beta$ asymmetry \cite{dubbers2019exotic}. This analysis along with the neutron lifetime anomaly has been discussed continuously in the literature (see e.g. Refs. \cite{markisch2019measurement,markisch2020accurate,tan2019laboratory,alonso2021strange,dubbers2021precise,berezhiani2021possible,broussard2021experimental,kuno2020precision,czarnecki2019radiative,giacosa2020measurement,elahi2020neutron,volya2020assessment}). Therefore, the neutron lifetime anomaly is still far from being fully conclusive. A specific dark decay channel: $n \rightarrow \chi e^+ e^-$ has been ruled out by the PERKEO II experiment \cite{klopf2019constraints}, where the limits on the corresponding branching ratios and the mass scales of dark matter particles has also been provided. A new measurement scheme, which can verify the explanation of neutron lifetime anomaly via neutron-mirror neutron oscillations, has also been designed \cite{broussard2021experimental}. To summarize, we consider that the neutron lifetime anomaly remains a puzzle and a reasonable theoretical explanation needs to be constructed.

In the formalism of the SM and its minor extensions, the manifestations of neutral particle oscillation (or mixing) can mainly be classified into two types: (\RomanNumeralCaps{1}) \textbf{Pure oscillation}; (\RomanNumeralCaps{2}) \textbf{Impure oscillation}. Generally, an oscillation process, no matter which type of oscillation it belongs to, is essentially contributed by the on-shell absorptive and off-shell dispersive mixing amplitudes. The classification of oscillation can be made based on whether an interference between oscillation and decay occurs or whether the new physics effects contribute to the absorptive mixing amplitude. The absorptive mixing amplitude is dominated by long-distance interactions mediated by on-shell intermediate states or decay products \cite{kagan2021dispersive}. Specifically, in the type-\RomanNumeralCaps{2} oscillation, there can be an interference between oscillation and decay, or else new physics effects may contribute to the absorptive mixing amplitude in a non-trivial way. For instance, the $K^0$-$\bar{K}^0$ oscillation (mixing) \cite{gell1955behavior,lee1957remarks,christenson1964evidence} can be classified into the type-\RomanNumeralCaps{2} oscillation, because both $K^0$ and $\bar{K}^0$ could decay into two or three Pions and their oscillation is accompanied by an interference between oscillation and decay \cite{sozzi2008discrete,bigi2009cp}. The origin of the $K^0$-$\bar{K}^0$ mixing effects can mainly be explained by weak interactions of the SM, where the size of the mixing can be calculated via e.g. loop-level diagrams (see e.g. Refs. \cite{sozzi2008discrete,bigi2009cp}). Other examples of the type-\RomanNumeralCaps{2} oscillation include $B^0$-$\bar{B}^0$ oscillation \cite{abashian2001measurement,aubert2001measurement,aubert2004direct,chao2005improved}, $D^0$-$\bar{D}^0$ oscillation \cite{aaij2019observation}, and etc. On the contrary, in the type-\RomanNumeralCaps{1} oscillation, the new physics effects do not contribute to the absorptive mixing amplitude and there is no interference between oscillation and decay. For instance, neutrino oscillation ($\nu_e \rightleftarrows \nu_\mu \rightleftarrows \nu_\tau$) \cite{fukuda1998evidence,fukuda2000tau,ahmad2001measurement,agafonova2010observation,an2012observation,abe2013evidence} can be classified into the type-\RomanNumeralCaps{1} oscillation. Unlike the type-\RomanNumeralCaps{2} oscillation, all the mixing angles in the type-\RomanNumeralCaps{1} oscillation are free parameters, which are introduced into the theories by a brute-force approach and can only be determined by experiments. For instance, in neutrino oscillation, the mixing angles and phases contained in the Pontecorvo–Maki–Nakagawa–Sakata (PMNS) matrix \cite{maki1962remarks} are free input parameters that can only be extracted from experiments and cannot be calculated from theories within the SM. The origin of the mixing angles in the type-\RomanNumeralCaps{1} oscillation remains a puzzle \cite{hagedorn2019lepton} and such angles may come from some short-distance interactions from new physics. The main differences between the two types of oscillation in the formalism of the SM are summarized in the Tab. \ref{appctab} in the Appendix \ref{appac}.

\begin{figure}[b] 
\centering
\includegraphics[scale=1.0,width=0.75\linewidth]{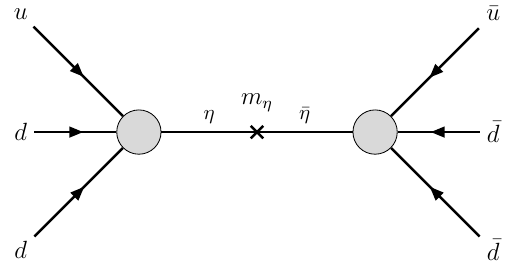}
\caption{The $n$-$\bar{n}$ oscillation can be induced at the tree level through the mixing between neutron and the elementary particle $\eta$ \cite{kalara1983supersymmetric,zwirner1983observable,mohapatra1986neutrino,aitken2017baryogenesis}. 
}
\label{netanbar}
\end{figure}

The hierarchy of the mixing angles and family masses can be explained by the partial compositeness model \cite{kaplan1991flavor}, where the Lagrangian can be divided into three sectors, such as the elementary, composite, and mixing sectors \cite{contino2007warped}. According to this model, both the composite and the elementary particles of the SM may not necessarily be mass eigenstates, whereas their superpositions could be mass eigensates \cite{contino2007warped}. The partial compositeness model predicts various new phenomena, such as proton-positron oscillation \cite{rajpoot1984proton}, neutron-neutrino oscillation \cite{rajpoot1984proton}, $\rho^0$-$\gamma$ oscillation \cite{nambu1962rare,gell1962decay,jegerlehner2011rho}, and etc. As a special case of the partial compositeness model, the mixing between neutron ($n$) and elementary neutral particle ($\eta$) violates $\mathcal{B}$ and $\mathcal{L}$ by one unit while conserving their difference $(\mathcal{B}-\mathcal{L})$, and may give rise to many interesting and observable consequences.

In what follows, we analyze the physical consequences arising from the $n$-$\eta$ mixing, such as the $n$-$\bar{n}$ oscillation, and discuss about the expected signal observability at the present and future experiments.

\section{Neutron-neutral particle mixing \label{nnmixing}}

The mixing between $n$ and $\eta$ can be mediated by color-multiplet scalar bosons, which can be contained in some new physics models based on higher symmetries such as $SU(4)_c \times SU(2)_L \times SU(2)_R$ \cite{pati1974lepton,pati1975erratum,mohapatra1980local}, $SU(5)$ \cite{georgi1974unity}, and etc. Such scalar bosons belong to a large family of particles including diquarks, leptoquarks, and etc. Among them, the diquarks can transform as a color triplet or a color sextet. For the sake of specificity, we assume that the diquarks transform as a color triplet but the conclusions remain valid for color sextet. The relevant operators can be written down in two different scenarios, which correspond to the representations $(3, 1, 2/3)$ and $(3, 1, -4/3)$ respectively under the SM group $SU(3)_c \times SU(2)_L \times U(1)_Y$\footnote{Here, the hypercharge is defined by $Y \equiv 2(Q-I_{3L})$.} (see e.g. Refs. \cite{babu2007unified,allahverdi2013natural,dev2015tev,aitken2017baryogenesis,jin2018nucleon}):
\begin{equation}
\begin{split}
\hat{O}_1 \equiv& \lambda_{ij} \epsilon^{\alpha \beta \gamma} \phi_{\alpha} \bar{u}_{i\beta} d^c_{j\gamma}  + \mu_{ij} \phi^{\dagger}_{\alpha} \bar{\eta}_i d^c_{j\alpha}  + \frac{1}{2}m_{\eta} \bar{\eta}_i \eta^c_i \\
&+ m_{\phi}^2 \phi^{\dagger}_{\alpha}\phi_{\alpha} + \text{H.c.}
\label{o1}
\end{split}
\end{equation}
\begin{equation}
\begin{split}
\hat{O}_2 \equiv& \lambda_{ij} \epsilon^{\alpha \beta \gamma} \phi_{\alpha} \bar{d}_{i\beta} d^c_{j\gamma} + \mu_{ij} \phi^{\dagger}_{\alpha} \bar{\eta}_i u^c_{j\alpha}  + \frac{1}{2}m_{\eta} \bar{\eta}_i\eta^c_i \\
&+ m_{\phi}^2 \phi^{\dagger}_{\alpha}\phi_{\alpha} + \text{H.c.}
\label{o12}
\end{split}
\end{equation}
Here, $\eta$ may have a non-zero lepton number ($\mathcal{L} \equiv 1$) and may intertwine with dark matter or else its decay products could be dark matter candidates \cite{allahverdi2013natural}. Furthermore, $\eta$ has to be neutral as required by the charge conservation law. $\alpha$, $\beta$, and $\gamma$ are the color indices. $\lambda_{ij}$ and $\mu_{ij}$ are two dimensionless coupling constants, where $i$ and $j$ are the generation indices. The $SU(4)_c$ symmetry limit requires that the coupling constants tend to be equal to each other ($\lambda \simeq \mu$) \cite{babu2009neutrino,babu2013post}. Since the discussions are only valid up to the order of the magnitude, the terms with permutations and the corresponding normalization factors are omitted for simplicity of notation. In addition, since we only focus on the first generation of quarks and leptons, the relevant coupling constants in our discussion are $\lambda_{11}$ and $\mu_{11}$ and the generation indices on $\eta$ can be suppressed. Currently, there is no direct experimental constraints on $\lambda_{11}$ and $\mu_{11}$, but instead there are some constraints on their combination with other coupling constants associated with different generations or flavors. The superscript $c$ denotes charge conjugation. $m_\eta$ is the mass of the elementary neutral particle ($\eta$). $m_\phi$ is the mass of the color multiplet bosons and thus it is associated with the new physics energy scale.

The flavor-changing neutral current (FCNC) effects can also be mediated by the color-multiplet scalar bosons \cite{mohapatra2008diquark,babu2009neutrino,saha2010constraining,dorvsner2011limits,barr2012observable,babu2013post,arnold2013phenomenology,babu2013expectations,fortes2013flavor,patra2014post,sahoo2015scalar,addazi2015exotic,dev2015tev,kim2019correlation,fridell2021probing}. The phenomenology of the FCNC effects mediated by diquarks and leptoquarks has been intensively studied \cite{mohapatra2008diquark,babu2009neutrino,saha2010constraining,dorvsner2011limits,barr2012observable,babu2013post,arnold2013phenomenology,babu2013expectations,fortes2013flavor,patra2014post,sahoo2015scalar,addazi2015exotic,dev2015tev,kim2019correlation,fridell2021probing}. The manifestation of the FCNC effects includes various processes such as meson-antimeson oscillation, rare decay modes of meson, lepton flavor violation (LFV), and etc. The measurements of such processes can provide a powerful tool to put severe constraints on the parameter space of new physics models. As an important feature, the derived bounds are not usually imposed on a single coupling constant but instead they are imposed on the product of the coupling constants with different flavors or generations, namely $|\lambda_{ij}\mu_{ij}|$, where the indices are not correlated. Furthermore, the derived bounds depend on the masses of the color multiplet bosons. For example, the derived bounds on the product of the coupling constants from the meson-antimeson oscillation processes roughly scale as the mass $m_{\phi}$ or the squared-mass $m_{\phi}^2$ \cite{babu2009neutrino,dorvsner2011limits,babu2013expectations,babu2013post,fortes2013flavor}. The estimated bounds in the literature vary remarkably and it is difficult to compare them as a result of the differences in the choice of theoretical models and experimental data. Quantitatively, if we are only interested in the appealing scenario where the masses of the color multiplet bosons (i.e. the new physics energy scales) are accessible to a direct detection at the LHC or future high-energy experiments (see e.g. Ref. \cite{hao2020connection}), which roughly corresponds to the range from several TeV to several $10$ TeV, the upper bounds on the product of the coupling constants $|\lambda_{ij}\mu_{ij}|$ can be more preferred to be restricted in the range from the order of $10^{-4}$ to the order of 1, roughly. For the sake of specificity, we choose some typical values, namely $|\lambda_{11} \mu_{11}| \simeq 10^{-1}$, $10^{-2}$, and $10^{-3}$, which are in general consistent with the FCNC constraints in our discussion. Smaller coupling constants tend to give rise to smaller masses of the color multiplet bosons. Although some calculated bounds on the coupling constants from the FCNC effects can be more restrictive than the typical values we choose, we could adjust the mass of $\eta$ so that our results are consistent with the FCNC constraints as well as the direct searches for the color multiplet bosons at the LHC as will be discussed below.

Similar to Eq. (\ref{o1}), the operators that give rise to proton decay can be allowed and a list of such operators can be found, e.g. in Ref. \cite{barr2012observable}. As a specific example, the color multiplet bosons that transform as $(3, 1, 2/3)$ or $(3, 1, 8/3)$ according to the SM symmetry can be present and may give rise to disastrously rapid proton decay. In order to forbid too rapid proton decay, new physics symmetries can be implemented \cite{gu2011baryogenesis,arnold2013phenomenology,dev2015tev}. In our case, we could assign a $Z_2$-odd quantum number to $\eta$, quark singlet ($u_R$, $d_R$), and doublet ($q_L$) of $SU(2)_L$, while assigning a $Z_2$-even quantum number to $\phi$, lepton singlet ($e_R$), and doublet ($l_L$) of $SU(2)_L$. In this way, the proton decay is forbidden by the $Z_2$-symmetry. Here, $q_{R/L}$ ($l_{R/L}$) stands for the right and left handed spinors which are defined by $q_{R/L} (l_{R/L}) \equiv P_{R/L} q (P_{R/L} l)$ with $P_{R/L}\equiv(1\pm \gamma^5)/2$.

Moreover, we could move a step back and consider the absence of the $Z_2$-symmetry. the color multiplet bosons that lead to the mixing between $n$ and $\eta$ transforms as $(3, 1, 2/3)$ or $(3, 1, -4/3)$ under the SM group $SU(3)_c \times SU(2)_L \times U(1)_Y$, while the color multiplet bosons that lead to the proton decay transforms as $(3, 1, 2/3)$ or $(3, 1, 8/3)$ under the same group. There could be a mass hierarchy between these two sets of color multiplet bosons. If the color multiplet bosons, which lead to proton decay, have a relatively heavier mass, the rate of proton decay could be slow enough and thus could be consistent with the present experimental limits.

When written down in terms of the neutron field, the relevant operators that account for the $n$-$\eta$ mixing in the absence of external (magnetic) fields can be given by (see e.g. Ref. \cite{mckeen2018neutron})
\begin{equation}
\hat{O}_3 \equiv \bar{\eta} i \cancel{\partial}\eta + \frac{1}{2}m_{\eta} \bar{\eta}  \eta^c + \bar{n} (i \cancel{\partial} -m_{n} ) n  + \delta \bar{n}\eta + \text{H.c.},
\label{oeq2}
\end{equation}
with \cite{jin2018nucleon}
\begin{equation}
\delta \equiv \frac{\lambda_{11}\mu_{11} \lvert \psi_q (0) \rvert^2 }{m_{\phi}^2}.
\end{equation}
Here, $m_n$ is the mass of neutron. The following substitutions  \cite{kaplan1991flavor,jin2018nucleon}: $udd \rightarrow \lvert \psi_q (0) \rvert^2 n$, $u^c d^c d^c \rightarrow \lvert \psi_q (0) \rvert^2 n^c$ have been made. $\psi_q (0)$ is the overlap factor of quarks. Lattice QCD calculations give the value $|\psi_q (0)|^2 = 0.0144 (3)(21)$ GeV$^3$ \cite{aoki2017improved}, where the numbers in the parentheses are the statistical and systematic uncertainties, respectively. We have assumed that antineutron has the same overlap factor of quarks as neutron does.

Without external magnetic fields, the mixing angle for the $n$-$\eta$ mixing can be the same as the one for the $\bar{n}$-$\bar{\eta}$ mixing. However, since the magnetic dipole moment of a neutron is oppositely oriented to the one of an anti-neutron \cite{phillips2016neutron}, the mixing angle for the $n$-$\eta$ mixing differs from the one for the $\bar{n}$-$\bar{\eta}$ mixing in the presence of the external magnetic fields. Since $\eta$ is a new particle outside the SM, it is quite natural to assume that the magnetic dipole moment of $\eta$ is negligible so that it barely interacts with electromagnetic fields. In this case, the effective mass matrix can be given by \cite{kaplan1991flavor}
\begin{align}
&\begin{aligned}
W&
=
\left[
\begin{array}{cc}
W_{11} & W_{12}\\
W_{21} & W_{22}
\end{array}
\right]\\
&
=
\left[
\begin{array}{lc}
M_{11}\mp \vert \mu_{n} B \vert  -\frac{i}{2}\Gamma_{11} &M_{12}-\frac{i}{2}\Gamma_{12}\\
M^{*}_{12}-\frac{i}{2}\Gamma^{*}_{12} &M_{22}-\frac{i}{2}\Gamma_{22}
\end{array}
\right],
\end{aligned}
\label{masseq6}
\end{align}
where $B$ is the external magnetic field and $\mu_{n}=g_n \mu_N/2$ is the magnetic dipole moment of neutron. Here, the neutron g-factor $g_n$ has the value: $g_n \simeq -3.826$ \cite{tiesinga2021codata} and the nuclear magneton has the value: $\mu_N \simeq 3.152 \times 10^{-8}$ eV$\cdot$T$^{-1}$ \cite{tiesinga2021codata}. In the ILL experiment, the magnetic field in the neutron propagation region can be as low as $B \lesssim 1 \times 10^{-8}$ T \cite{baldo1994new} and to be conservative we can choose the maximum value: $B=1 \times 10^{-8}$ T in our analysis. In Eq. (\ref{masseq6}), the minus and plus sign corresponds to the $n$-$\eta$ and $\bar{n}$-$\bar{\eta}$ mixing, respectively. The off-diagonal matrix elements $M_{12}$ and $\Gamma_{12}$ describe the dispersive and absorptive mixing amplitudes of the effective mass matrix, respectively.

In the presence of the $n$-$\eta$ mixing, the mass eigenstates $\ket{n_1}$ and $\ket{n_2}$ can be expressed as linear superposition of the interaction eigenstates $\ket{n}$ and $\ket{\eta}$: 
\begin{align}
&\begin{aligned}
\ket{n_1} &\equiv c_1 \ket{n} + \epsilon_1 \ket{\eta},
\label{n1n21}
\end{aligned}\\
&\begin{aligned}
\ket{n_2} &\equiv \epsilon_2 \ket{n} + c_2 \ket{\eta}.
\label{n1n22}
\end{aligned}
\end{align}
Here, $c_{1,2}$ and $\epsilon_{1,2}$ represent the mixing coefficients, which are the elements of the transformation matrix $T$ that is used to diagonalize the effective mass matrix $W$ (see Appendix \ref{appa} for more details).

In the presence of the external magnetic fields, the transformation matrix associated with the $n$-$\eta$ mixing ($T_1$) is different from the one associated with the $\bar{n}$-$\bar{\eta}$ mixing ($T_2$):
\begin{align}
T_{1,2}&
=
\left[
\begin{array}{ccl}
\cos\theta_{1,2} &\sin\theta_{1,2}\\
-\sin\theta_{1,2} &\cos\theta_{1,2}
\end{array}
\label{t1eq}
\right].
\end{align}
Here, $\theta_1$ and $\theta_2$ are the mixing angles associated with the $n$-$\eta$ mixing and the $\bar{n}$-$\bar{\eta}$ mixing, respectively, and they satisfy the following expressions:
\begin{align}
&\begin{aligned}
\theta_{1,2} &\equiv \arctan\Big(\frac{2\delta }{m_n \mp \vert \mu_{n} B \vert -m_{\eta}+ \Delta W_{1,2} }\Big),
\label{tan1}
\end{aligned}
\end{align}
with
\begin{equation}
\Delta  W_{1,2}  \equiv \Big[ (m_n \mp \vert \mu_{n} B \vert -m_{\eta})^2+4\delta^2 \Big]^{\frac{1}{2}}.
\end{equation}

Based on Eq. (\ref{t1eq}), the probability of the $n$-$\eta$ ($\eta$-$n$) and $\bar{n}$-$\bar{\eta}$ ($\bar{\eta}$-$\bar{n}$) oscillations in the same external magnetic field can be respectively given by \cite{mohapatra2009neutron,cohen2019quantum,babu2022theoretical}
\begin{align}
&\begin{aligned}
P_{n \rightarrow \eta},P_{\bar{n} \rightarrow \bar{\eta} } &=  \frac{4 \delta^2 \sin^2\Big[\frac{\sqrt{(m_n \mp \vert \mu_{n} B \vert -m_{\eta})^2 + 4 \delta^2 }}{2} t \Big]}{(m_n \mp \vert \mu_{n} B \vert -m_{\eta})^2 + 4 \delta^2}\\
&= \sin^2{(2\theta_{1,2})} \sin^2\Big(\frac{\phi_{1,2}}{2}\Big), 
\label{eq18}
\end{aligned}
\end{align}
where the CP-even phases $\phi_1$ and $\phi_2$ are defined by
\begin{equation}
\phi_{1,2} \equiv \Big[ (m_n \mp \vert \mu_{n} B \vert -m_{\eta})^2 + 4 \delta^2 \Big]^{\frac{1}{2}} t. 
\end{equation}
As can be seen, without external magnetic fields, the mixing angles, the CP-even phases, the transformation matrices, and the oscillation probabilities in Eq. (\ref{eq18}) take the same value for the particle and anti-particle sectors, i.e. $\theta \equiv \theta_1 = \theta_2$, $\phi \equiv \phi_1 = \phi_2$, $T \equiv T_1 = T_2$, $P_{n \rightarrow \eta} = P_{\bar{\eta} \rightarrow \bar{n}}$.

The time evolution of the interaction eigenstates can be given by
\begin{equation}
\begin{split}
\ket{n(t)} =& \frac{1}{c^2+\epsilon^2}\Big[ \Big( c^2 e^{-i \omega_{1}t-\Gamma_{1}t} + \epsilon^2 e^{-i \omega_{2}t-\Gamma_{2}t}\Big) \ket{n} \\
&+ \Big( c \epsilon e^{-i \omega_{1}t-\Gamma_{1}t} - c \epsilon e^{-i \omega_{2}t-\Gamma_{2}t} \Big) \ket{\eta} \Big],    
\end{split}
\end{equation}
\begin{equation}
\begin{split}
\ket{\eta(t)} =& \frac{1}{c^2+\epsilon^2}\Big[ \Big( c \epsilon e^{-i \omega_{1}t-\Gamma_{1}t} - c \epsilon e^{-i \omega_{2}t-\Gamma_{2}t} \Big) \ket{n} \\
&+ \Big( \epsilon^2 e^{-i \omega_{1}t-\Gamma_{1}t} + c^2 e^{-i \omega_{2}t-\Gamma_{2}t}\Big) \ket{\eta} \Big].    
\end{split}
\end{equation}
Here, $\omega_{1,2}$ and $\Gamma_{1,2}$ are the masses and widths of the mass eigenstates $\ket{n_1}$ and $\ket{n_2}$, respectively. As can be seen, a neutron that is created at the beginning ($t=0$) can later be detected to be an $\eta$ particle with a specific probability. The above-mentioned $n$-$\eta$ mixing may lead to many interesting and observable consequences, one of which is the $n$-$\bar{n}$ oscillation.

\subsection{$n$-$\bar{n}$ oscillation}

Fig. \ref{netanbar} shows that the $n$-$\bar{n}$ oscillation can be achieved at the tree level indirectly through elementary neutral particles (e.g. $\eta$ in our case) as intermediate states, instead of through composite particles (e.g. mirror neutron $n^\prime$ \cite{mohapatra2005some,berezhiani2006neutron,berezhiani2016neutron,berezhiani2021possible}). In this case, an elementary particle (e.g. $\eta$) with a non-zero lepton number ($\mathcal{L}=1$) enjoys some advantages over a composite particle (e.g. $n^{\prime}$ \cite{mohapatra2005some,berezhiani2006neutron,berezhiani2016neutron,berezhiani2021possible}) with a non-zero baryon number ($\mathcal{B}=1$). To begin with, many grand unified models based on higher symmetry groups, such as $SU(5)$ \cite{georgi1974unity} or $SO(10)$ \cite{fritzsch1975unified}, are featured by a unified description of quarks and leptons. The $n$-$\bar{n}$ oscillation process mediated by $\eta$ can serve as a promising probes for such grand unified models. In addition, the number of new particles introduced into the model is relatively smaller in the scenario where the $n$-$\bar{n}$ oscillation is mediated by elementary particles. Moreover, the $n$-$\bar{n}$ oscillation mediated by the composite particle (e.g. $n^\prime$ \cite{mohapatra2005some,berezhiani2006neutron,berezhiani2016neutron,berezhiani2021possible}) can be described by dimension-9 operators, which are highly suppressed by the new physics energy scale, while the $n$-$\bar{n}$ oscillation mediated by the elementary particle $\eta$ can be described by dimension-6 operators \cite{dev2015tev}, which are less suppressed and thus more natural.

In previous studies (see e.g. Refs \cite{allahverdi2013natural,allahverdi2014kev,jin2018nucleon,fajfer2021colored,mckeen2020long}), the following restriction is imposed on $m_{\eta}$ to make the proton decay kinetically forbidden:
\begin{equation}
m_p -m_e \lesssim m_{\eta} \lesssim m_p +m_e.
\end{equation}
Here, $m_p$ and $m_e$ are the proton and electron mass respectively. An even more stringent restriction $m_{\eta}> 937.9$ MeV can be derived from the stability of $^9$Be \cite{mckeen2016c,mckeen2018neutron}. However, such restrictions are very stringent and lack of experimental support. Another possibility that has not been excluded is that $m_{\eta}$ may lie outside this narrow range while the stability of proton and nuclei (e.g. $^9$Be) can be guaranteed by imposing additional assumptions or symmetries \cite{babu2007unified,allahverdi2013natural,dev2015tev,mckeen2016c}. In this work, we loosen the restriction on $m_{\eta}$ to the whole range where the decay of neutron into $\eta$ is kinetically allowed:
\begin{equation}
m_{\eta} \lesssim m_n.
\end{equation}
Here, the masses $m_n$ and $m_{\eta}$ definitely refer to the masses of the mass eigenstates $\ket{n_1}$ and $\ket{n_2}$.

In what follows, we analyze the physical consequences (i.e. the $n$-$\bar{n}$ oscillation) arising from the $n$-$\eta$ mixing and discuss about their expected signal observability at the present and future experiments. We focus on two different scenarios, depending on whether the $n$-$\eta$ mixing contributes to the absorptive mixing amplitude. In the scenario where the neutral particle oscillation is type-\RomanNumeralCaps{1}, we evaluate the $n$-$\bar{n}$ oscillation probability and analyze the observability of the Majorana phase and CP-violating effects. In the scenario where the neutral particle oscillation is type-\RomanNumeralCaps{2}, we analyze the testable implications on the masses and lifetimes of the mass eigenstates that are resulted from the $n$-$\bar{n}$ oscillation mediated by $\eta$. In both scenarios, the lower limits imposed by the results of the searches for $n$-$\bar{n}$ oscillations on the masses of the color multiplet bosons can be estimated. Finally, we discuss about the observability of the geometric phase associated with the $n$-$\eta$ mixing and comment on its possible measurement scheme.

\subsection{Majorana phase, CP-violation, and oscillation probability \label{sec4}}

In this subsection, we focus on the type-\RomanNumeralCaps{1} oscillation, where the $n$-$\eta$ mixing contributes to the absorptive mixing amplitude. We evaluate the $n$-$\bar{n}$ oscillation probability and analyze the observability of the Majorana phase and the CP-violating effects. Furthermore, we derive the lower limits imposed by the results of the searches for $n$-$\bar{n}$ oscillations on the masses of the color multiplet bosons.

The connection between Majorana phases and CP-violation has been studied intensively in the neutrino sector \cite{bilenky1980oscillations,schechter1980neutrino,schechter1981neutrino,doi1981cp}. In this case, the Majorana phases may have non-trivial impact on the observable effects, such as the decay rate of the neutrino-less double $\beta$ decay \cite{doi1981cp}, the antineutrino-neutrino oscillations \cite{bilenky1980oscillations,schechter1980neutrino,schechter1981neutrino,doi1981cp,fritzsch2001describe}, and etc. If neutrinos are Majorana-type particles, there could be more CP-violating phases, possibly leading to more observable consequences \cite{doi1981cp}. However, in neutrino-neutrino oscillations, the Majorana phase may not have observable effects since it is difficult to produce a coherent state for different types of neutrinos, which does not correlate to the charged leptons \cite{giunti2010no}. Even though the Majorana phases can lead to non-trivial effects on the decay rate of the neutrino-less double $\beta$ decay \cite{doi1981cp}, it does not necessarily manifested as CP-violation \cite{de2003manifest}.

The $n$-$\bar{n}$ oscillation process can be considered as a potential probe for CP-violation based on CPT and Lorentz symmetries \cite{berezhiani2015neutron}. However, such a process may not necessarily lead to observable CP-violating effects \cite{fujikawa2015neutron}. It was pointed out that CP-violation cannot be observable in the $n$-$\bar{n}$ oscillation unless there is an interaction or an interference between amplitudes \cite{berezhiani2019neutron}. Although a single neutron is not a Majorana particle, if there is a direct mixing between neutron and antineutron, their superposition could be a Majorana particle \cite{fujikawa2016parity}. In this case, since the CP-violating phase in the $n$-$\bar{n}$ mixing matrix can be absorbed into the definition of the neutron field \cite{fujikawa2016parity}, there is no observable CP-violating effect. Nevertheless, the situation could be different, if a neutron mixes with a Majorana particle $\eta$.

We assume that the neutral elementary particle $\eta$ satisfies the following Majorana condition \cite{schechter1981neutrino}: 
\begin{equation}
\eta^c \equiv \eta e^{i \xi},
\label{majorana}
\end{equation}
where $\xi$ is the Majorana phase, which comes from the Majorana nature of $\eta$ and cannot be eliminated by a field redefinition (i.e. a rephrasing transformation). In what follows, we will discuss the observable consequences of the Majorana phase $\exp(i \xi)$ contained in the $n$-$\eta$ mixing matrix and reveal its CP-violating nature.

Without loss of generality, in the following discussions we assume the initial neutron is right handed. We can read from Eq. (\ref{oeq2}) that the relevant operators responsible for the $n$-$\eta$ mixing take the following form \cite{de2003manifest}:
\begin{equation}
\begin{split}
\hat{O} &=\frac{\lambda_{11}\mu_{11} |\psi_q (0)|^2}{m_{\phi}^2}\bar{n}_{R} \eta_L + \text{H.c.}\\
&= -\frac{\lambda_{11}\mu_{11} |\psi_q (0)|^2 e^{-i\xi}}{m_{\phi}^2}\bar{\eta}_{R} n^c_L + \text{H.c.},
\end{split}
\end{equation}
where in the last step we have used the condition in Eq. (\ref{majorana}). In the following discussions, we assume that the external magnetic fields are absent and thus the mixing angles $\theta_1$ and $\theta_2$ take the same value, i.e. $\theta \equiv \theta_1 = \theta_2$, $T \equiv T_1 = T_2$. The results can be generalized to the case where the external magnetic fields are present, simply by substituting $\theta_1$ for $\theta$ in the $n$-$\eta$ mixing and by substituting $\theta_2$ for $\theta$ in the $\bar{n}$-$\bar{\eta}$ mixing. The phase factor $\exp(i \xi)$ can be arranged into the off-diagonal elements of the transformation matrix $T$ \cite{schechter1981neutrino}:
\begin{equation}
T = 
\left[
\begin{array}{cc}
\cos\theta & e^{i\xi} \sin\theta\\
-e^{-i\xi}\sin\theta & \cos\theta
\end{array}
\right].
\end{equation}
Note the phase factors can be equivalently moved to the elements in the second column of the matrix $T$   without triggering additional measurable effects \cite{giunti2010no}.

Previous studies have shown that the CP-violating effects induced by the Majorana phase may be observable in the neutrino-antineutrino (or antineutrino-neutrino) oscillations \cite{schechter1981neutrino,doi1981cp,fritzsch2001describe,de2003manifest}. Possible measurement schemes for the Majorana phase have been proposed for the antineutrino-neutrino ($\bar{\nu}$-$\nu$) oscillations  \cite{schechter1981neutrino,fritzsch2001describe}. As an example, such a measurement can be accomplished by an oscillation process and two sequences of scattering processes \cite{fritzsch2001describe}: (1) $\mu^+ n \rightarrow \bar{\nu}_\mu p$, (2) $\bar{\nu}_\mu \rightarrow \nu_\mu$, (3) $\nu_\mu n \rightarrow \mu^- p$. Here, CP-violation can be achieved through the $\bar{\nu}_\mu$-$\nu_\mu$ oscillation process. Similar to the $\bar{\nu}_\mu$-$\nu_\mu$ oscillation process, CP-violation can also be achieved through the $\eta$-$\bar{\eta}$ oscillation process contained in the entire $n$-$\bar{n}$ oscillation process. In this case, the $n$-$\bar{n}$ oscillation process induced by the intermediate state $\eta$ can be divided into three sub-processes: (1) $n_R$-$\eta_R$ oscillation, (2) $\eta_R$-$\bar{\eta}_L$ oscillation, (3) $\bar{\eta}_L$-$\bar{n}_L$ oscillation. Note in the second sub-process, there is a chirality flip due to the Majorana mass term. The probability of the entire process can be expressed as
\begin{equation}
P_{n \rightarrow \bar{n}}\equiv P_{n_R \rightarrow \eta_R} P_{\eta_R \rightarrow \bar{\eta}_L} P_{\bar{\eta}_L \rightarrow \bar{n}_L}.
\label{ppp1}
\end{equation}

The $\eta_R$-$\bar{\eta}_L$ oscillation amplitude can be given in the same manner as described in the antineutrino-neutrino oscillation case \cite{schechter1981neutrino,doi1981cp,fritzsch2001describe,de2003manifest}:
\begin{equation}
A_{\eta_R \rightarrow \bar{\eta}_L} = \sum_{i=1}^2 \Big[ K_1 m_i \Big(T_{\eta i}\Big)^2 e^{-i \omega_i t} \Big].
\label{eq47}
\end{equation}
Here, $m_i$ are the masses of $n$ and $\eta$, respectively. $K_1$ is a kinetic factor and up to a trivial phase factor satisfies the expression: $K_1 = 1/\omega_1$ \cite{schechter1981neutrino,fritzsch2001describe,de2003manifest}, where $\omega_1$ is the energy of the initial neutron. $T_{\eta i}$ represents the elements in the second row of the matrix $T$. The $\eta_R$-$\bar{\eta}_L$ oscillation probability is then given by \cite{schechter1981neutrino,fritzsch2001describe,de2003manifest}
\begin{equation}
\begin{split}
P_{\eta_R \rightarrow \bar{\eta}_L}=& \lvert K_1 \rvert^2 \Big[m_\eta^2 \cos^4\theta + m_n^2 \sin^4\theta \\
             &+ \frac{1}{2} m_n m_\eta  \sin^2(2\theta) \cos(\phi + 2\xi)\Big].
\end{split}
\label{pnnbar1}
\end{equation}
Here, we have assumed that the external magnetic fields are absent and thus the CP-even phase $\phi_1$ satisfies the condition: $\phi \equiv \phi_1(B=0) \simeq \sqrt{(m_n-m_{\eta})^2 + 4 \delta^2} t$.

\begin{figure}[t] 
\centering
\includegraphics[scale=1.0,width=0.99\linewidth]{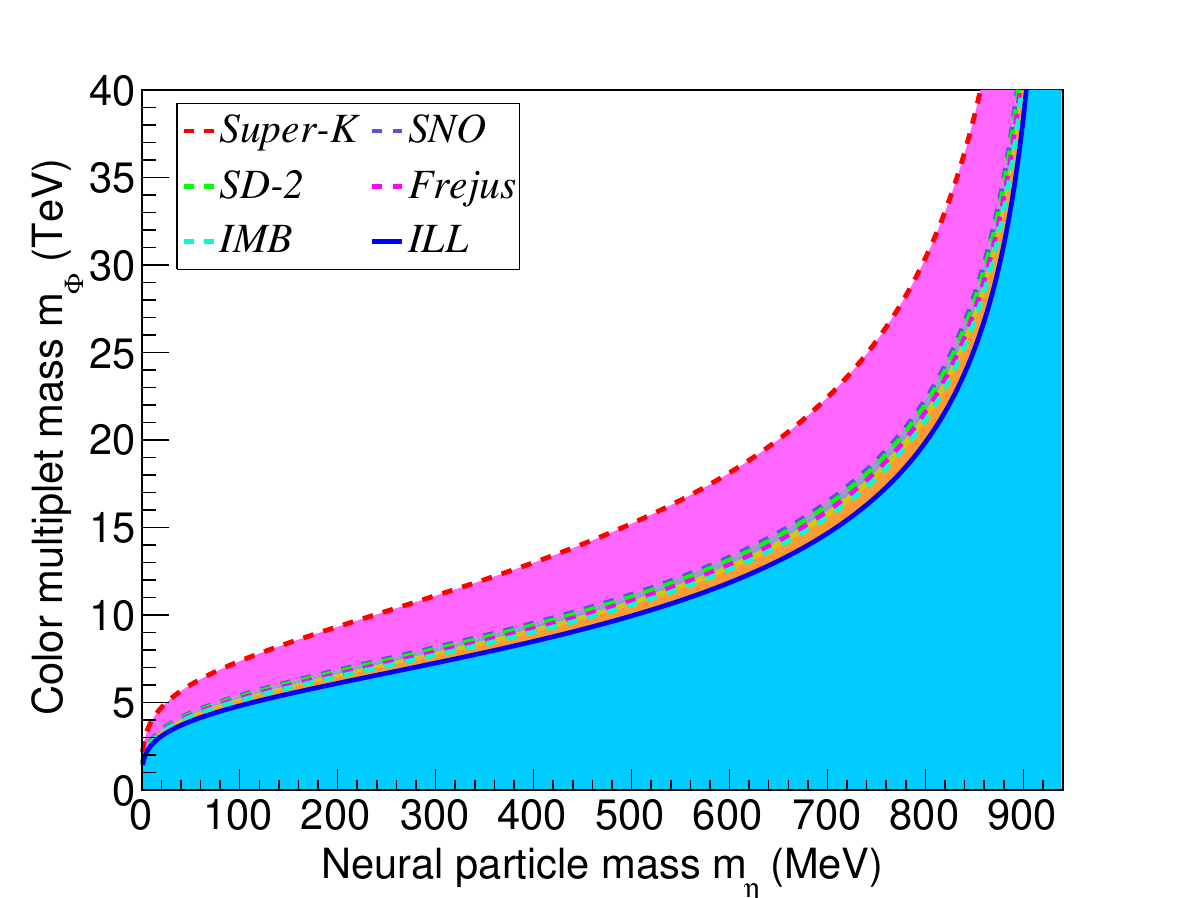}
\caption{The constraints imposed by the experimental searches for the $n$-$\bar{n}$ oscillation on the mass of the color multiplet boson $m_{\phi}$ as a function of $m_{\eta}$ with the coupling constants $|\lambda_{11}\mu_{11}| \equiv 10^{-1}$ in the framework of pure oscillation. The shaded regions have been excluded. (Color online)
}
\label{figratio5}
\end{figure}

\begin{figure}[t] 
\centering
\includegraphics[scale=1.0,width=0.99\linewidth]{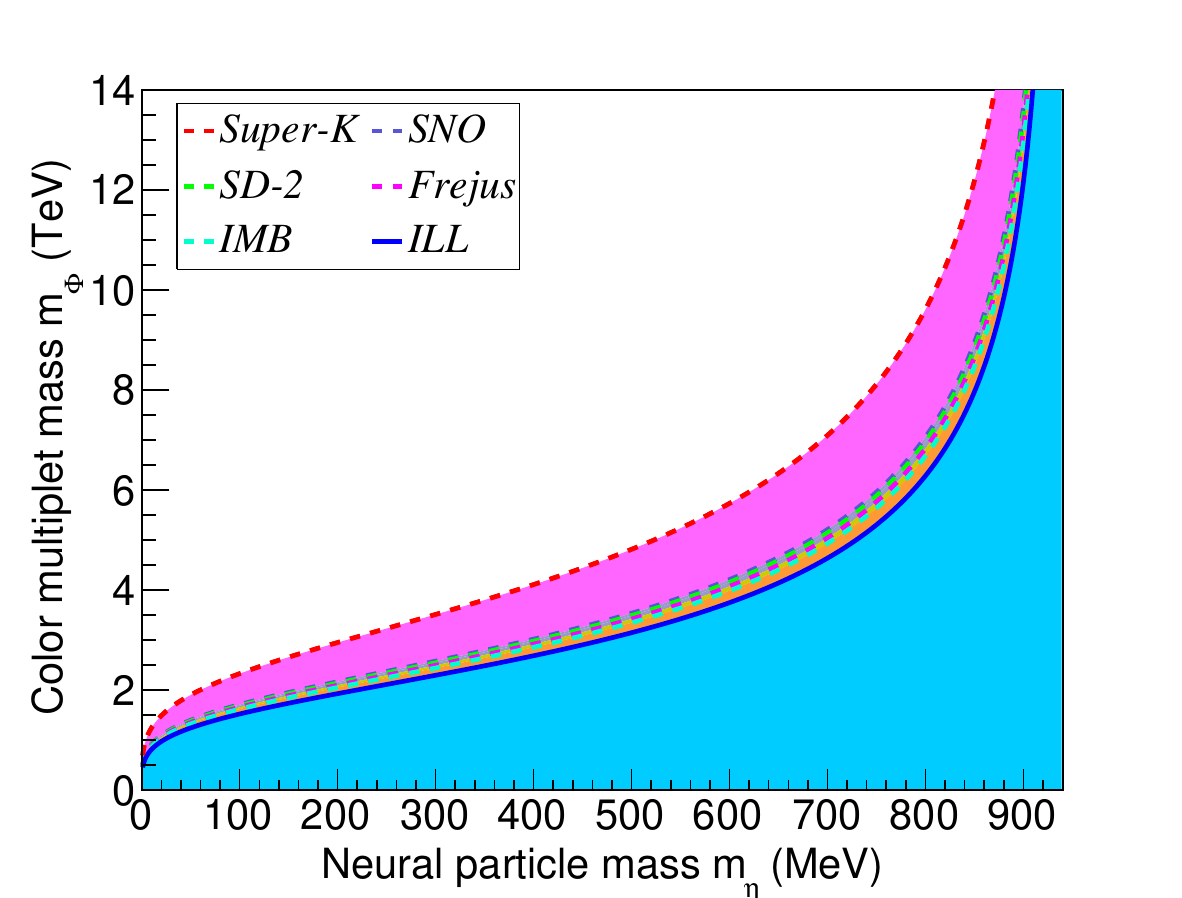}
\caption{The constraints imposed by the experimental searches for the $n$-$\bar{n}$ oscillation on the mass of the color multiplet boson $m_{\phi}$ as a function of $m_{\eta}$ with the coupling constants $|\lambda_{11}\mu_{11}| \equiv 10^{-2}$ in the framework of pure oscillation. The shaded regions have been excluded. (Color online)
}
\label{figratio6}
\end{figure}

Analogously, the antineutron-neutron ($\bar{n}$-$n$) oscillation, which is the CP-conjugate process of the $n$-$\bar{n}$ oscillation, can also be divided into three distinct sub-processes: (1) $\bar{n}_L$-$\bar{\eta}_L$ oscillation, (2) $\bar{\eta}_L$-$\eta_R$ oscillation, (3) $\eta_R$-$n_R$ oscillation. The corresponding probability of the entire process can be written as
\begin{equation}
P_{\bar{n} \rightarrow n}\equiv P_{\bar{n}_L \rightarrow \bar{\eta}_L} P_{\bar{\eta}_L \rightarrow \eta_R} P_{\eta_R \rightarrow n_R}.
\label{ppp2}
\end{equation}
Here, the $\bar{\eta}_L$-$\eta_R$ oscillation amplitude can be given by \cite{schechter1981neutrino,doi1981cp,fritzsch2001describe,de2003manifest}
\begin{equation}
A_{\bar{\eta}_L \rightarrow \eta_R} = \sum_{i=1}^2 \Big[ K_2 m_i \Big(T_{\eta i}^{*}\Big)^2 e^{-i \omega_i t} \Big].
\end{equation}
$K_2$ is another kinetic factor, which is different from $K_1$ by an irrelevant phase factor \cite{de2003manifest} and thus has the same modulus as $K_1$, i.e. $|K_1|=|K_2|$. 
The $\bar{\eta}_L$-$\eta_R$ oscillation probability is then given by \cite{schechter1981neutrino,fritzsch2001describe,de2003manifest}
\begin{equation}
\begin{split}
P_{\bar{\eta}_L \rightarrow \eta_R}=& \lvert K_2 \rvert^2 \Big[m_{n}^2 \sin^4\theta + m_{\eta}^2 \cos^4\theta  \\
             &+ \frac{1}{2} m_{n} m_{\eta} \sin^2(2\theta) \cos(\phi - 2\xi )\Big].
\end{split}
\label{pnnbar2}
\end{equation}
Here, we have also assumed that the external magnetic fields are absent and thus the CP-even phase $\phi_2$ satisfies the condition: $\phi \equiv \phi_2(B=0) \simeq \sqrt{(m_n-m_{\eta})^2 + 4 \delta^2} t$ too. Furthermore, in the absence of external magnetic fields, the following relations can be obtained from Eq. (\ref{eq18}):
\begin{equation}
\begin{split}
&P_{n_R \rightarrow \eta_R} = P_{\eta_R \rightarrow n_R} = P_{\bar{n}_L \rightarrow \bar{\eta}_L} = P_{\bar{\eta}_L \rightarrow \bar{n}_L}\\
=&\sin^2{(2\theta)} \sin^2\Big(\frac{\phi}{2}\Big).   
\end{split}
\label{eq46}
\end{equation}
In the non-relativistic scenario, i.e. $\omega_1 \simeq m_n$, Eq. (\ref{pnnbar1}) and (\ref{pnnbar2}) can be rewritten as \cite{schechter1981neutrino,fritzsch2001describe,de2003manifest}
\begin{equation}
\begin{split}
P_{\eta_R \rightarrow \bar{\eta}_L}=& \Big[\Big(\frac{m_\eta}{m_n}\Big)^2 \cos^4\theta + \sin^4\theta \\
             &+ \frac{1}{2} \Big( \frac{m_\eta}{m_n} \Big)  \sin^2(2\theta) \cos(\phi + 2\xi )\Big],
\end{split}
\label{pnnbar3}
\end{equation}
\begin{equation}
\begin{split}
P_{\bar{\eta}_L \rightarrow \eta_R}=& \Big[\Big(\frac{m_\eta}{m_n}\Big)^2 \cos^4\theta + \sin^4\theta \\
             &+ \frac{1}{2} \Big( \frac{m_\eta}{m_n} \Big)  \sin^2(2\theta) \cos(\phi - 2\xi)\Big].
\end{split}
\label{pnnbar31}
\end{equation}
Since the $\eta_R$-$\bar{\eta}_L$ and $\bar{\eta}_L$-$\eta_R$ oscillations are characterized by a chirality flip, the corresponding probabilities are suppressed by the mass of $\eta$ as expected. Based on Eq. (\ref{pnnbar3}) and (\ref{pnnbar31}), the probabilities for the $n$-$\bar{n}$ and $\bar{n}$-$n$ oscillation are respectively given by
\begin{equation}
\begin{split}
P_{n \rightarrow \bar{n}}^{B=0} =& \sin^4{(2\theta)} \sin^4\Big(\frac{\phi}{2}\Big) \Big[\Big(\frac{m_\eta}{m_n}\Big)^2 \cos^4\theta + \sin^4\theta \\
&+ \frac{1}{2} \Big( \frac{m_\eta}{m_n} \Big)  \sin^2(2\theta) \cos(\phi + 2\xi )\Big],
\end{split}
\label{pnnbar4}
\end{equation}
\begin{equation}
\begin{split}
P_{\bar{n} \rightarrow n}^{B=0}=& \sin^4{(2\theta)} \sin^4\Big(\frac{\phi}{2}\Big) \Big[\Big(\frac{m_\eta}{m_n}\Big)^2 \cos^4\theta + \sin^4\theta \\
&+ \frac{1}{2} \Big( \frac{m_\eta}{m_n} \Big)  \sin^2(2\theta) \cos(\phi - 2\xi )\Big].
\end{split}
\label{pnnbar6}
\end{equation}

\begin{widetext}
\begin{equation}
\begin{split}
A_{\text{CP}}(B=0) &\equiv \frac{P_{n \rightarrow \bar{n}} - P_{\bar{n} \rightarrow n}}{P_{n \rightarrow \bar{n}} + P_{\bar{n} \rightarrow n}}
=\frac{m_n m_{\eta}\sin^2{(2\theta)} [\cos(\phi + 2\xi)-\cos(\phi - 2\xi)]}{4m_n^2\sin^4{\theta} + 4m_{\eta}^2\cos^4{\theta} + m_n m_{\eta}\sin^2{(2\theta)} [\cos(\phi + 2\xi)+\cos(\phi - 2\xi)]}.
\end{split}
\label{cpeq50}
\end{equation}
\begin{equation}
\begin{split}
P_{n \rightarrow \bar{n}}^{B \neq 0} \simeq & \frac{1}{8} \Big[ \frac{4 \delta^2}{(m_n-|\mu_n B|-m_{\eta})^2+ 4\delta^2} \Big] \Big\{\Big( \frac{m_{\eta}}{m_n} \Big)^2 \Big[\frac{(m_n-|\mu_n B|-m_{\eta})^2 +2 \delta^2}{(m_n-|\mu_n B|-m_{\eta})^2+ 4\delta^2} + \sqrt{ \frac{(m_n-|\mu_n B|-m_{\eta})^2}{(m_n-|\mu_n B|-m_{\eta})^2+ 4\delta^2} } \Big] \\
&+ \Big[\frac{(m_n-|\mu_n B|-m_{\eta})^2+2 \delta^2}{(m_n-|\mu_n B|-m_{\eta})^2+ 4\delta^2} - \sqrt{ \frac{(m_n-|\mu_n B|-m_{\eta})^2}{(m_n-|\mu_n B|-m_{\eta})^2+ 4\delta^2} }\Big] \Big\} \Big[ \frac{4 \delta^2}{(m_n+|\mu_n B|-m_{\eta})^2+ 4\delta^2} \Big].    
\end{split}
\label{pnnmag}
\end{equation}
\end{widetext}

Similar to the CP-violating conditions given in Ref. \cite{de2003manifest}, Eq. (\ref{pnnbar4}) and (\ref{pnnbar6}) show that, if $\cos(\phi + 2\xi) \neq \cos(\phi - 2\xi)$ (i.e. $\phi \neq n \pi$ and $\xi \neq n \pi/2$, $n \in \mathbb{N}$), a CP-violating effect ($P_{n \rightarrow \bar{n}} \neq P_{\bar{n} \rightarrow n}$) due to the Majorana phase can be observable. Since the $n$-$\eta$ mixing does not contribute to the absorptive mixing amplitude, such manifestation of CP-violation is resulted from the phase mismatch between the dispersive and absorptive mixing amplitudes. This is different from the situation in the type-\RomanNumeralCaps{2} oscillation, where the Majorana phases of the dispersive and absorptive parts cancel out and there will be no CP-violation unless some conditions are satisfied \cite{mckeen2016c}. Eq. (\ref{cpeq50}) gives the corresponding CP asymmetry $A_{\text{CP}}(B=0)$ and indicates explicitly that a non-trivial CP-violation due to the Majorana phase can be expected in the $n$-$\bar{n}$ oscillation. Furthermore, the absolute value of the CP asymmetry $|A_{\text{CP}}|$ has the following maximum at the points $\phi= \pm \pi/2$ and $\xi= \pm \pi/4$, where the signs are not correlated:
\begin{equation}
\lvert A_{\text{CP}}^{\text{max}}(B=0)\rvert = \frac{m_n m_{\eta} \sin^2(2\theta)}{2 m_n^2 \sin^4{\theta}+ 2m_{\eta}^2\cos^4{\theta}}.
\end{equation}

$\mathcal{B}$-violation and CP-violation (along with C-symmetry violation) are two of the three conditions presented by Sakharov to explain the observed matter-antimatter asymmetry in our Universe \cite{sakharov1967violation}. The $n$-$\bar{n}$ oscillation accompanied by observable CP-violating effects provides an appealing scenario for exploring new physics effects and may open a promising avenue for explaining the origin of the matter-antimatter asymmetry.

At the beginning of Sec. \ref{nnmixing}, we have explained the reason why the restriction imposed on $m_{\eta}$ can be  chosen to be $m_{\eta} \lesssim m_n$. In the ILL experiment, the measurement of the $n$-$\bar{n}$ oscillation time was performed with a mean propagation time of neutron $\tau_m \equiv t \simeq 0.1$ s \cite{baldo1994new}. Nearly over the whole range [$m_{\eta} \in (0, m_n$)], the CP-even phase $\phi$ satisfies the condition: 
\begin{equation}
\phi = \sqrt{(m_n-m_{\eta})^2 + 4 \delta } \tau_m \gtrsim |m_n - m_{\eta}| \tau_m\gg 1.
\label{notquasi}
\end{equation}
Here, the CP-even phase $\phi$ is associated with the $n$-$\eta$ oscillation subprocess rather than the entire $n$-$\bar{n}$ oscillation process. The above condition holds even in the presence of external magnetic fields as the magnetic interaction term $|\mu_n B|\lesssim 6 \times 10^{-22}$ MeV is very small. Furthermore, this condition does not necessarily contradict the quasi-free condition $\Delta E t \ll 1$ given in Refs. \cite{mohapatra1980phenomenology,baldo1994new}, where the CP-even phase $\Delta E t$ is associated with the entire $n$-$\bar{n}$ oscillation process. A rough estimation shows that even in the absence of external magnetic fields, the condition $\phi \ll 1$ implies that $|m_n - m_{\eta}| \ll 6.6 \times 10^{-23}$ MeV. Since $n$ and $\eta$ are completely different particles, it is extremely unnatural to require that they have almost equal masses, i.e. $|m_n - m_{\eta}| \ll 6.6 \times 10^{-23}$ MeV. Therefore, in the following discussions, we employ the condition given in Eq. (\ref{notquasi}). In this limit, Eq. (\ref{pnnbar4}) takes the following form:
\begin{equation}
\begin{split}
&P_{n \rightarrow \bar{n}}^{B=0}\\
\simeq &\frac{1}{8} \Big[ \frac{4 \delta^2}{(m_n-m_{\eta})^2+ 4\delta^2} \Big]^2  \Big\{ \Big( \frac{m_{\eta}}{m_n} \Big)^2 \Big[\frac{(m_n-m_{\eta})^2 + 2\delta^2}{(m_n-m_{\eta})^2+ 4\delta^2} \\
&+ \sqrt{ \frac{(m_n-m_{\eta})^2}{(m_n-m_{\eta})^2+ 4\delta^2} } \Big] +\frac{(m_n-m_{\eta})^2 + 2\delta^2}{(m_n-m_{\eta})^2+ 4\delta^2}\\
& - \sqrt{ \frac{(m_n-m_{\eta})^2}{(m_n-m_{\eta})^2+ 4\delta^2} }\Big\}.
\end{split}
\label{longeq60}
\end{equation}

In Eq. (\ref{pnnbar4}) and (\ref{pnnbar6}), we have assumed that the external magnetic field is absent. In reality, the external magnetic fields may not be fully shielded and an unavoidable background magnetic field needs to be considered. Under the condition expressed in Eq. (\ref{notquasi}), the $n$-$\bar{n}$ oscillation probability in the presence of external magnetic fields can be given by Eq. (\ref{pnnmag}). In the Appendix \ref{appb}, the probabilities for the $n$-$\bar{n}$ and $\bar{n}$-$n$ oscillations without making the approximation associated with the limit $\phi_{1,2} \gg 1$ can be given by Eq. (\ref{fullpnn}) and (\ref{fullpnn2}), respectively.

\begin{table*}[t]
\caption{Results of the searches for $n$-$\bar{n}$ oscillations and the corresponding estimated oscillation probabilities. }
\begin{ruledtabular}
\begin{tabular}{l|ccccccc}                                         
\diagbox{Param.}{Exp.}& ILL \cite{baldo1994new}& IMB \cite{jones1984search} &KM \cite{takita1986search}&Frejus \cite{berger1990search}&SD-2 \cite{chung2002search}&SNO \cite{aharmim2017search}&Super-K \cite{abe2021neutron} \\\hline           
Candidates $S_{0}$ & 0 & 0 &0   &0 &5        &23       &11     \\
$\tau_{n\bar{n}}$ in matter (yr) & $-$ & $2.4 \times 10^{31}$ & $4.3 \times 10^{31}$ & $6.5 \times 10^{31}$ &  $7.2 \times 10^{31}$  & $3.0 \times 10^{31}$  & $3.6 \times 10^{32}$\\
Suppression $R$ (s$^{-1}$) &   $-$  & $1.0 \times 10^{23}$  & $1.0 \times 10^{23}$ & $1.4 \times 10^{23}$ & $1.4 \times 10^{23}$  & $2.5 \times 10^{22}$ & $5.17 \times 10^{22}$\\
$\tau_{n\bar{n}}$ in vacuum (s) & $8.6 \times 10^7$& $1.1 \times 10^8$ &  $1.2 \times 10^8$  & $1.2 \times 10^8$ &  $1.3 \times 10^8$   & $1.37 \times 10^8$  & $4.7 \times 10^8$\\ 
Probability $P_{n \rightarrow \bar{n}}$ & $1.4 \times 10^{-18}$ & $^a$$8.3\times 10^{-19}$ &$^a$$6.9\times 10^{-19}$ &$^a$$6.9\times 10^{-19}$ &$^a$$5.9\times 10^{-19}$ &$^a$$5.3\times 10^{-19}$ &$^a$$4.5\times 10^{-20}$\\
\end{tabular}
\end{ruledtabular}
\begin{tablenotes}
\footnotesize
\centering
\item[\emph{a}]{
$^a$ The oscillation probabilities are converted into the field-free vacuum values based on the oscillation times in vacuum.
}
\end{tablenotes}
\centering
\label{tabtwo}
\end{table*}

In the ILL experiment, the measurement of the $n$-$\bar{n}$ oscillation was carried out using cold neutrons with a beam intensity of around $1.25 \times 10^{11}$ neutrons per second and a neutron propagation time of around $0.1$ s \cite{baldo1994new}. In the quasi-free condition, the $n$-$\bar{n}$ oscillation probability can be estimated by the expression: $P_{n \rightarrow \bar{n}} \simeq t^2/\tau^2_{n\bar{n}}$ \cite{baldo1994new}, where $\tau_{n\bar{n}}$ is the $n$-$\bar{n}$ oscillation time in field-free vacuum. Under the assumption that the typical propagation time of free neutrons is $0.1$ s, the $n$-$\bar{n}$ oscillation times can be translated into the $n$-$\bar{n}$ oscillation probabilities. The results of the searches for $n$-$\bar{n}$ oscillations have been reported by various experiments and the corresponding estimated oscillation probabilities are shown in Tab. \ref{tabtwo}.
As an example, the lower bound on the $n$-$\bar{n}$ oscillation time reported by the ILL experiment is about $8.6 \times 10^7$ s \cite{baldo1994new}, which, approximately, corresponds to the oscillation probability of the order of $10^{-18}$ (see e.g. Ref. \cite{berezhiani2021possible}).

Based on Eq. (\ref{pnnmag}), the constraints on the masses of the color multiplet boson $m_{\phi}$ can be estimated. As discussed earlier, we choose some typical values for the product of the coupling constants. Fig. \ref{figratio5} and \ref{figratio6} show the constraints imposed by the experimental searches for the $n$-$\bar{n}$ oscillation on the masses of the color multiplet boson $m_{\phi}$, corresponding to the coupling constants $|\lambda_{11}\mu_{11}| \simeq 10^{-1}$, $10^{-2}$ respectively. As can be seen from Fig. \ref{figratio5} and \ref{figratio6}, comparing with the scenario where the coupling constant is $|\lambda_{11}\mu_{11}| \simeq 10^{-1}$, the scenario with $|\lambda_{11}\mu_{11}| \simeq 10^{-2}$ predicts a smaller mass of color multiplet boson, i.e. a smaller new physics energy scale. Furthermore, the  constraints on the mass of color multiplet boson varies gently from $1$ to $8$ TeV ($|\lambda_{11}\mu_{11}| \simeq 10^{-2}$) and from $5$ to $25$ TeV ($|\lambda_{11}\mu_{11}| \simeq 10^{-1}$) throughout the entire range of the allowed $m_{\eta}$ values except in the vicinity of the neutron mass $m_n$. This illustrates that the bounds on the new physics energy scale are in general insensitive to $m_{\eta}$ unless $m_{\eta}$ lies within the vicinity of the neutron mass. In the vicinity of the neutron mass $m_n$, the derived bounds on the color multiplet bosons can increase rapidly with respect to the mass of $\eta$. if we choose smaller coupling constants which also satisfies some more restrictive FCNC constraints, we would obtain less competitive bounds on the mass of the color multiplet bosons, i.e. the bounds are even smaller than the ones imposed by the LHC experiments. In this case, we could adjust the mass of $\eta$ so that the derived bounds on the mass of the color multiplet bosons are consistent with the limits imposed by the direct searches at the LHC. This means that the mass of $\eta$ cannot be too far away from the neutron mass. To summarize, in order to satisfy all the constraints the mass of $\eta$ cannot be randomly chosen and we need to consider a balance between these constraints.

\subsection{Mass and lifetime\label{masslife}}

\begin{figure}[b] 
\centering
\includegraphics[scale=1.0,width=0.75\linewidth]{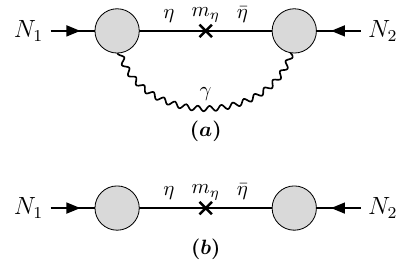}
\caption{The possible contributions to the $n$-$\bar{n}$ oscillation \cite{mckeen2016c}: (a) The absorptive amplitude $\Gamma_{12}$ is mainly originated from the process mediated by the on-shell $\gamma$ and $\eta$; (b) The dispersive amplitude $M_{12}$ is mainly originated from the process mediated by the off-shell $\eta$ \cite{mckeen2016c}. 
}
\label{m12}
\end{figure}

In this subsection, we focus on the type-\RomanNumeralCaps{2} oscillation, where the $n$-$\eta$ mixing not only contributes to the off-shell dispersive mixing amplitude ($M_{12}$) but also contributes to the on-shell absorptive mixing amplitude ($\Gamma_{12}$).

We analyze the $n$-$\bar{n}$ oscillation originated from the $n$-$\eta$ mixing and estimate its implications on masses and lifetimes. In this case, the $n$-$\bar{n}$ oscillation occurs indirectly through the  dispersive and absorptive amplitudes as depicted in Fig. \ref{m12} \cite{mckeen2016c}. Within the framework of the type-\RomanNumeralCaps{2} oscillation, the possibility of establishing CP-violation in baryon oscillations has been discussed in Ref. \cite{mckeen2016c}. In this case, since the Majorana phase contained in the $n$-$\eta$ mixing contributes equally to the dispersive and absorptive mixing amplitudes, the phases from these two parts tend to cancel out and there is no observable CP-violating effect. The lower bounds on the masses of the color multiplet bosons can be derived by employing the results of the searches for $n$-$\bar{n}$ oscillations. Furthermore, we analyze the compatibility between the interpretation of the neutron lifetime anomaly and the interpretation of the $n$-$\bar{n}$ oscillation with regard to the $n$-$\eta$ mixing.

The $n$-$\eta$ mixing can lead to the $n$-$\bar{n}$ oscillation, which can subsequently give rise to a mismatch between the neutron and antineutron interaction eigenstates and their mass eigenstates. Due to the $n$-$\bar{n}$ oscillation, the linear superposition of the neutron and antineutron interaction eigenstates gives rise to mass eigenstates:
\begin{align}
&\begin{aligned}
\ket{N_1} &\equiv c^{\prime}_1 \ket{n} + \epsilon^{\prime}_1 \ket{\bar{n}},
\end{aligned}\\
&\begin{aligned}
\ket{N_2} &\equiv \epsilon^{\prime}_2 \ket{n} + c^{\prime}_2 \ket{\bar{n}}.
\end{aligned}
\end{align}
Here, $\ket{N_1}$ and $\ket{N_2}$ are the two mass eigenstates arising from the $n$-$\bar{n}$ oscillation and, according to Eq. (\ref{eq6}) in Appendix \ref{appa}, their mass difference satisfies the condition: $|m_{N_1}-m_{N_2}| \lesssim 2 |M_{12}|$. $c^{\prime}_{1,2}$ and $\epsilon^{\prime}_{1,2}$ are the corresponding mixing coefficients associated with the entire $n$-$\bar{n}$ oscillation process and they, in general, are different from the ones presented in Eq. (\ref{n1n21}) and (\ref{n1n22}) associated with the $n$-$\eta$ oscillation process. Due to the condition: $c^{\prime}_{1,2} \gg \epsilon^{\prime}_{1,2}$, the $\ket{N_1}$ state is predominantly composed of the $\ket{n}$ state while the $\ket{N_2}$ state is predominantly composed of the $\ket{\bar{n}}$ state.

Fig. \ref{m12} (a) shows the main possible contribution to the absorptive amplitude, which is mainly originated from the process mediated by the on-shell $\gamma$ and $\eta$ \cite{mckeen2016c}. Fig. \ref{m12} (b) shows that the dispersive amplitude $M_{12}$ is mainly arising from the process mediated by the off-shell $\eta$ \cite{mckeen2016c}. The two processes can be described by the effective Lagrangian \cite{mckeen2016c,mckeen2018neutron,fornal2018dark}:
\begin{equation}
\begin{split}
\mathscr{L}_{\text{eff}} \equiv& \bar{\eta} (i \cancel{\partial} -m_{\eta}) \eta + \delta(\bar{n}\eta + \text{H.c.})\\ 
& + \bar{n} \Big(i \cancel{\partial} -m_{n} + \frac{g_n}{2m_n} \sigma_{\mu \nu} F^{\mu \nu}\Big) n.
\end{split}
\end{equation}
Although there is no direct coupling between the neutral particle $\eta$ and the vector field $F^{\mu \nu}$, the following Lagrangian, which is responsible for the process depicted in the Fig. \ref{m12} (a), can be obtained by the diagonalization of the mass matrix \cite{mckeen2016c,mckeen2018neutron,fornal2018dark}:
\begin{equation}
\mathscr{L}_{\text{eff}} \supset \frac{g_n \sin\theta}{2m_n}  \bar{n}  \sigma_{\mu \nu} F^{\mu \nu} \eta + \text{H.c.}
\end{equation}
The mixing angle $\theta$, which is associated with the $n$-$\eta$ mixing, can be obtained from Eq. (\ref{tan1}) \cite{mckeen2018neutron,fornal2018dark}:
\begin{equation}
\theta \simeq \frac{\lambda_{11}\mu_{11} |\psi_q (0)|^2}{m_{\phi}^2 (m_n - m_{\eta})},
\end{equation}

Following Ref. \cite{mckeen2016c}, the absorptive amplitude $\Gamma_{12}$, which describes the $n$-$\bar{n}$ oscillation through on-shell intermediate states, can be approximately given by 
\begin{equation}
\Gamma_{12} \simeq \frac{g_n^2 \lambda_{11}^2 \mu_{11}^2 |\psi_q (0)|^4 m_{\eta}}{64 \pi m_{\phi}^4  (m_n - m_{\eta})^2} \Big(1-\frac{m_\eta^2}{m_n^2} \Big)^3.
\label{gamma12}
\end{equation}
Note Eq. (\ref{gamma12}) differs from Eq. (11) of Ref. \cite{mckeen2016c} by a factor associated with the $n$-$\eta$ mixing. The dispersive amplitude $M_{12}$, which describes the $n$-$\bar{n}$ transition through off-shell intermediate particles, can be estimated according to Fig. \ref{m12} (b). Assuming that the single-particle process makes a dominating contribution, the general form of $M_{12}$ can be given by (see e.g. Refs. \cite{marshak1969theory,mckeen2016c})
\begin{equation}
\begin{split}
M_{12} &\simeq \frac{1}{2 \sqrt{s}} \sum_{i} \frac{|A(n \rightarrow \psi_i)|^2}{s-m_i^2} \\
       &\simeq \frac{\lambda_{11}^2 \mu_{11}^2 |\psi_q (0)|^4 m_\eta}{m_{\phi}^4 (m_n^2 - m_{\eta}^2)}.
\end{split}
\label{m12eq}
\end{equation}
In the first step, the sum runs over all the intermediate particles $\psi_i$. The Mandelstam variable $s$ is defined in the conventional way and in the rest frame of the neutron it takes the value $s=m_n^2$. $A(n \rightarrow \psi_i) \equiv \bra{\eta(p^{\prime})} \hat{O} \ket{n(p)}$ is the amplitude in connection with the $n$-$\eta$ oscillation. In the second step, an approximation is made based on the assumption that $M_{12}$ is predominantly contributed by the process depicted in Fig. \ref{m12} (b), where $\eta$ is the only intermediate particle. According to Eq. (\ref{gamma12}) and (\ref{m12eq}), the ratio between $\Gamma_{12}$ and $M_{12}$ takes the form:
\begin{equation}
\kappa \equiv \frac{\Gamma_{12}}{M_{12}} = \frac{g_n^2 (m_n - m_{\eta})^2 (m_n + m_{\eta})^4}{64 \pi m_n^6}.
\label{eqkappa}
\end{equation}
This expression shows that by taking the ratio between $\Gamma_{12}$ and $M_{12}$ a large degree of uncertainty arising from the parameters, such as $\lambda_{11}$, $\mu_{11}$, $m_{\phi}$ and $|\psi_q (0)|^2$, could be eliminated. 

Once $\Gamma_{12}$ and $M_{12}$ are known, the observable consequences arising from the $n$-$\bar{n}$ oscillation can be obtained based on Eq. (\ref{eqkappa}). The $n$-$\bar{n}$ oscillation time reported by the ILL experiments is around $0.86 \times 10^{8}$ s \cite{baldo1994new}, which imposes a stringent constraint: $M_{12} \lesssim |\delta| \lesssim 7.7 \times 10^{-30}$ MeV \cite{nussinov2020using,berezhiani2021possible}. With the help of this constraint, the impact of the $n$-$\bar{n}$ oscillation on the masses and lifetimes of the mass eigenstates $N_1$ and $N_2$ can be evaluated based on Eq. (\ref{eq6}) and (\ref{eq7}) in Appendix \ref{appa}. Furthermore, the bounds on the mass of the color multiplet boson can also be estimated according to Eq. (\ref{m12eq}).

\begin{figure}[t] 
\centering
\includegraphics[scale=1.0,width=0.99\linewidth]{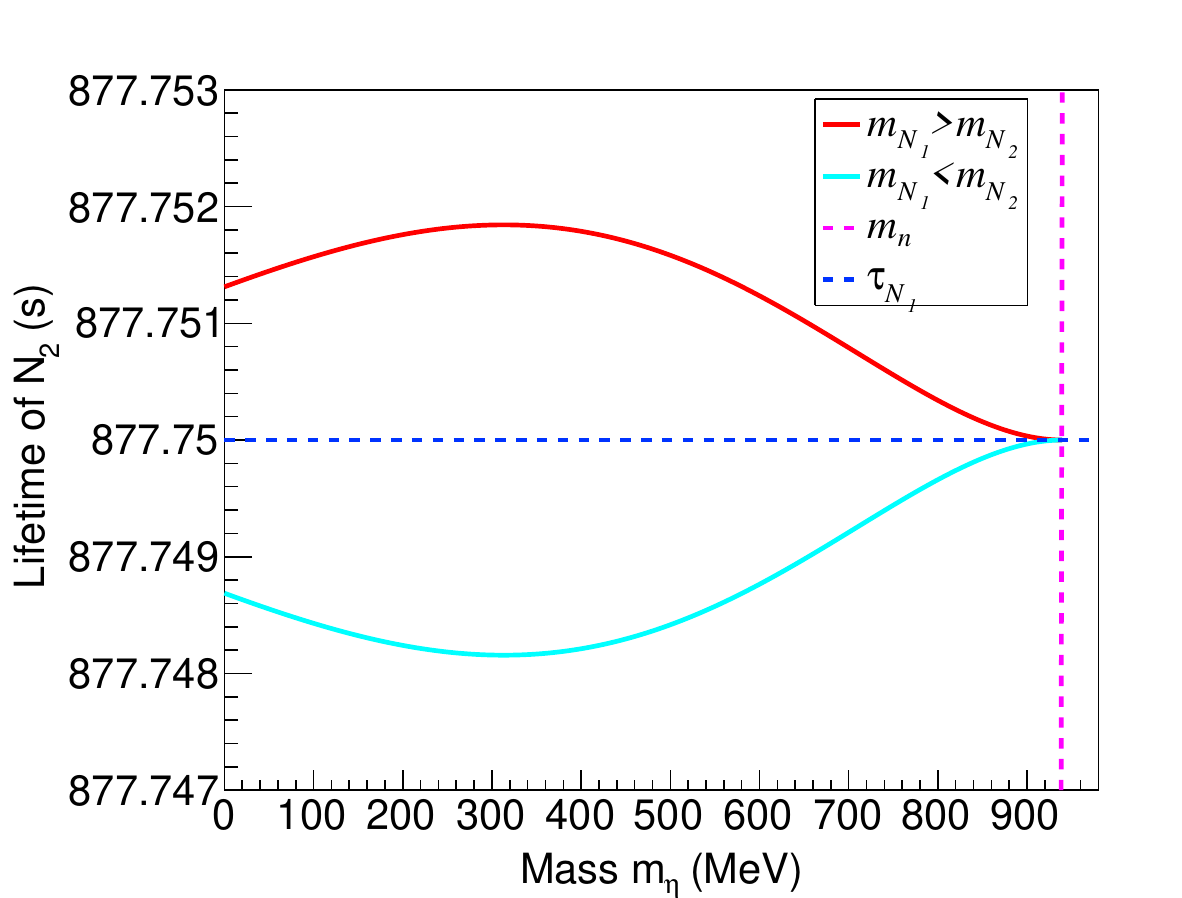}
\caption{The predicted lifetime of the mass eigenstate $N_2$ based on the results of the $n$-$\bar{n}$ oscillation experiments in the frame work of impure oscillation. (Color online) 
}
\label{figratio1}
\end{figure}

\begin{figure}[t] 
\centering
\includegraphics[scale=1.0,width=0.99\linewidth]{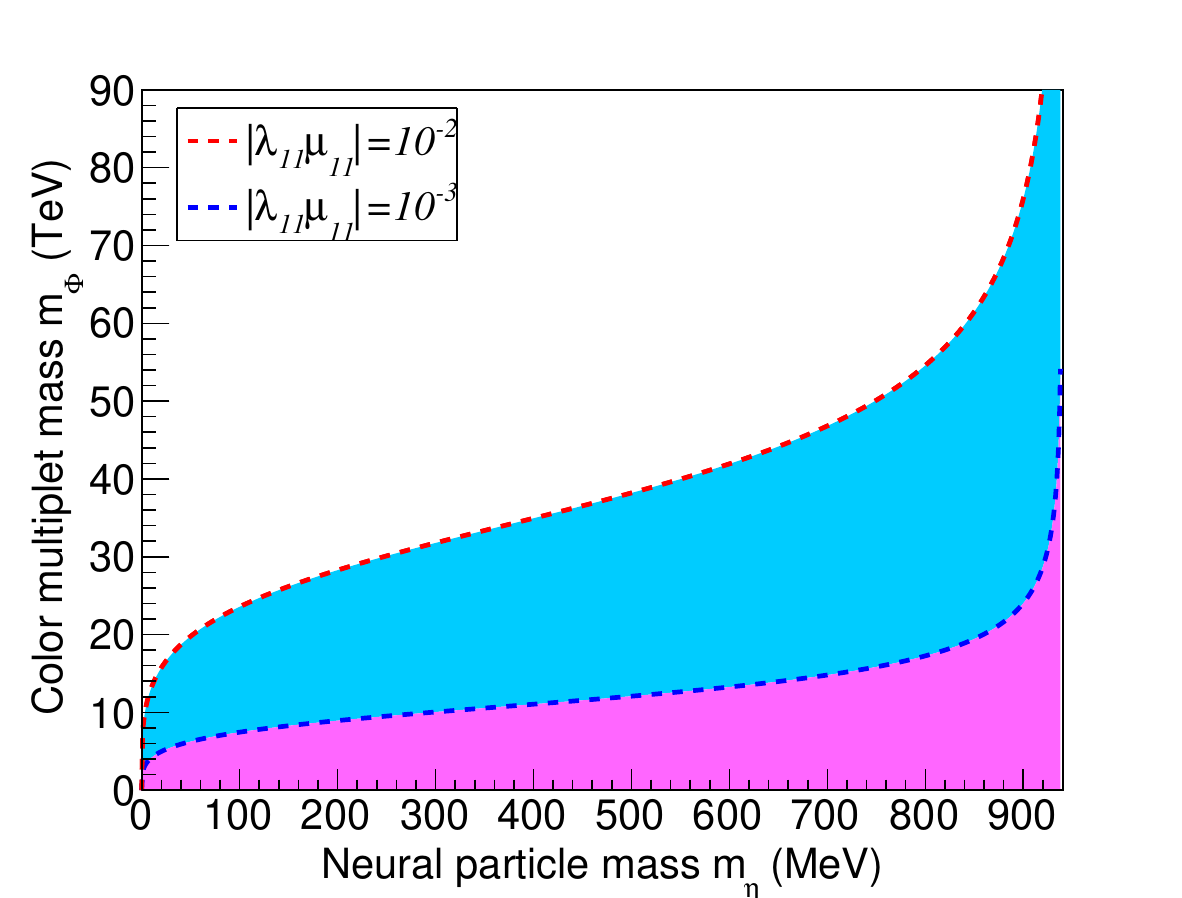}
\caption{The constraints on the mass of the color multiplet boson $m_{\phi}$ as a function of $m_{\eta}$ in the framework of impure oscillation. The shaded regions have been excluded. (Color online)
}
\label{figratio2}
\end{figure}

Fig. \ref{figratio1} shows the predicted lifetime of the mass eigenstate $N_2$ based on the experimental searches for the $n$-$\bar{n}$ oscillation. The solid curve in red corresponds to the predicted lifetime of $N_2$ in the scenario where $N_1$ is heavier than $N_2$. The solid curve in blue corresponds to the predicted lifetime of $N_2$ in the scenario where $N_1$ is lighter than $N_2$. The horizontal dashed line is the experimental lifetime of $N_1$ reported by the trap experiment \cite{gonzalez2021improved}. As can be seen from Fig. \ref{figratio1}, the difference in the lifetime between $N_1$ and $N_2$ has a maximum value ($\Delta \tau \simeq 1.84 \times 10^{-3}$ s) around the point $m_{\eta} = m_n/3$. This implies that when measuring the lifetimes of $N_1$ (mainly composed of $n$) and $N_2$ (mainly composed of $\bar{n}$), a lifetime difference as large as $1.84 \times 10^{-3}$ s would be expected. Such a lifetime difference is probably beyond the reach of the present experiments but may lie within the detectable regions in future experiments.

Fig. \ref{figratio2} shows the constraints on the mass of the color multiplet boson ($m_{\phi}$) in the framework of the type-\RomanNumeralCaps{2} oscillation. The regions below the solid curves have been excluded. Similarly, if we are only interested in the appealing scenario where the masses of the color multiplet boson lie within the range from several TeV to several $10$ TeV, the coupling constants $|\lambda_{11} \mu_{11}|$ are more favorable to select the values of $10^{-2}$ and $10^{-3}$. As can be seen from Fig. \ref{figratio2}, comparing with the scenario where the coupling constant is $|\lambda_{11} \mu_{11}| \simeq 10^{-2}$, the scenario with $|\lambda_{11} \mu_{11}| \simeq 10^{-3}$ predicts a smaller new physics energy scale. On the one hand, the bounds on the masses of the color multiplet boson vary gently from several TeV to several $10$ TeV throughout the entire range of the allowed $m_{\eta}$ values except in the vicinity of the neutron mass $m_n$. Similar trends have also been found in subsection \ref{sec4}. If we require that $m_{\phi}$ lies within the range, which is accessible to a direct detection at the LHC or future high-energy experiments \cite{allahverdi2013natural,deppisch2015neutrinos}, the mass of the neutral particle $m_{\eta}$ cannot be too close to the neutron mass $m_n$. On the other hand, however, if the mass of $\eta$ is located in the vicinity of the neutron mass $m_n$, the derived bounds on the mass of the color multiplet bosons can increase rapidly. As mentioned earlier, some calculated bounds on the coupling constants arising from the FCNC measurements can be smaller than the typical values we choose. Smaller coupling constants can loosen the bounds on the mass of the color multiplet bosons and thus make our bounds less competitive. Nevertheless, we could adjust the mass of $\eta$ so that the derived bounds on the mass of the color multiplet bosons are consistent with the experimental limits imposed by the direct searches at the LHC. This suggests that the mass of $\eta$ cannot be too far away from the neutron mass.

The neutron lifetime anomaly, which refers to the discrepancy in the measured neutron lifetime between two different experimental approaches, has attracted great attention recently (see e.g Ref. \cite{zyla2020review}). This discrepancy suggests that the branching fraction for the decay of neutron into proton through the $\beta$-decay is around $99\%$ and thus the invisible branching fraction is around $\Gamma_{a}/\Gamma_{n} \simeq 0.01$ \cite{fornal2018dark,mckeen2018neutron,dubbers2019exotic,fajfer2021colored}, where $\Gamma_{n}$ and $\Gamma_{a}$ are the neutron decay rate and the anomaly-induced decay rate, respectively. As discussed in Sec. \ref{sec1}, the neutron lifetime anomaly remains a puzzle and a reasonable theoretical explanation needs to be constructed.

Concerning the neutron lifetime anomaly, it is necessary to figure out what the manifestations of a neutron state really are in its production, interaction, and detection processes. Pariticularly, in the detection process, it is important to distinguish between what particles have been created and what particles have been really detected in experiments \cite{griffiths2008introduction}. If we maintain that a pure particle should have a definite mass and a definite lifetime, only the mass eigenstate can be treated as pure particles because it has a well-defined mass and lifetime. For example, in the $K^0$-$\bar{K}^0$ mixing \cite{gell1955behavior,lee1957remarks,christenson1964evidence}, the mass eigenstates $K_L$ and $K_S$, which are the mixtures of the $K^0$ and $\bar{K}^0$ mesons, can be treated as pure particles because $K_L$ and $K_S$ have well-defined masses and lifetimes, but $K^0$ and $\bar{K}^0$ cannot   according to this definition.

In the SM, a commonly recognized neutron state $\ket{n}$, which almost inclusively decays into electron, proton, and antineutrino through the $\beta$-decay process ($n \rightarrow pe^{-}\bar{\nu}_e$) \cite{zyla2020review}, is mainly created by the weak and strong interactions and may not necessarily coincide with a mass eigenstate. It is analogous to the explanation for the solar neutrino problem \cite{davis1968search,ahmad2001measurement}, where neutrinos are produced and detected in weak interaction eigenstates rather than in mass eigenstates. In this manner, the commonly recognized neutron state would not have a well-defined lifetime and the neutron lifetime discrepancy, which lies between the disappearance of the neutrons in a trap (bottle) and the detection of neutron in a beam, may be resolved in a simple way. The trap and bottle experiments are performed through the detection of the neutron disappearance \cite{gonzalez2021improved}, where the measurements are associated with the mass eigenstate. Alternatively, the beam experiments are performed through the detection of the $\beta$-decay products, such as proton and electron \cite{czarnecki2018neutron}, where the measurements are associated with the weak interaction eigenstate. Therefore, what have been detected in the trap (bottle) and beam experiments are completely two different things and may correspond to different manifestations of neutron states. 

In the presence of the $n$-$\eta$ mixing, a neutron that is created at the beginning ($t=0$) can later be detected as an $\eta$ particle with a specific probability. If we assume that $m_{\eta}$ satisfies the condition: $|m_n - m_{\eta}| \gg 6.6 \times 10^{-23}$ MeV, the CP-even phase in Eq. (\ref{eq18}) would satisfy the condition: $\phi_{1,2} \gg 1$ and the probability that a neutron transits into an $\eta$ particle at later time when it propagates through space can be approximated as
\begin{equation}
\frac{\Gamma_{a}}{\Gamma_{n}} \equiv P_{n \rightarrow \eta}  \simeq  \frac{2 \delta^2}{(m_n - |\mu_n B| -m_{\eta})^2 + 4 \delta^2}.
\label{pnnapprox}
\end{equation}
With this expression, the mass of the color multiplet boson can be estimated by
\begin{equation}
m_{\phi} \simeq \Big(\frac{2\Gamma_n-4\Gamma_{a}}{\Gamma_{a}} \Big)^{\frac{1}{4}}  \Big[ \frac{\lambda_{11} \mu_{11} |\psi_q(0)|^2}{m_n- |\mu_n B|-m_{\eta}} \Big]^{\frac{1}{2}}.
\end{equation}
Similarly, we assume that the coupling constants take the typical value $|\lambda_{11} \mu_{11}|\simeq 10^{-2}$. If we require that the mass of the color multiplet boson lies within the experimentally interesting range at the LHC or future high energy experiments, namely $1 \lesssim m_{\phi} \lesssim 10$ TeV, the mass difference should satisfy the condition: $2.0 \times 10^{-2} \lesssim |m_{n}-m_{\eta}|\lesssim 2.0$ MeV, which is automatically consistent with the condition: $|m_n - m_{\eta}| \gg 6.6 \times 10^{-23}$ MeV and thus justifies the approximation in Eq. (\ref{pnnapprox}).

Next, we analyze the compatibility between the interpretation of the neutron lifetime anomaly and the interpretation of the $n$-$\bar{n}$ oscillation experiments in connection with the $n$-$\eta$ mixing. The $n$-$\bar{n}$ oscillation probability is given by $P_{n \rightarrow \bar{n}}\equiv P_{n \rightarrow \eta} P_{\eta \rightarrow \bar{\eta}} P_{\bar{\eta} \rightarrow \bar{n}}$ [see Eq. (\ref{ppp1}) in subsection \ref{sec4}]. Here, the chirality subscripts are omitted. The observability of the color multiplet boson at the LHC or future high energy experiments requires that the mass difference $|m_n - m_{\eta}|$ cannot be too large. In this case, the $\eta$-$\bar{\eta}$ oscillation probability approximately takes the value $P_{\eta \rightarrow \bar{\eta}} \simeq 1$ according to Eq. (\ref{pnnbar3}). The lower bound imposed by the ILL experiment on the $n$-$\bar{n}$ oscillation probability is roughly in the order of $10^{-18}$ (see e.g. Ref. \cite{berezhiani2021possible}), which is much smaller than the $n$-$\eta$ oscillation probability defined in Eq. (\ref{pnnapprox}) and seems inconsistent with the interpretation of the neutron lifetime anomaly using the $n$-$\eta$ mixing. This inconsistency can be resolved by assuming that the neutral particle $\eta$ has a much shorter lifetime comparing with the neutron. In the $n$-$\bar{n}$ oscillation experiments, a small fraction of the neutrons, which are created at the beginning from the neutron source, can convert into the $\eta$ particles with a small probability as they propagate through space. Most of the $\eta$ particles would decay rapidly into invisible products before they oscillate into neutrons and thus only a small fraction of the $\eta$ particles could oscillate into neutrons. According to this assumption, the $n$-$\bar{n}$ oscillation probability can be rewritten as $P_{n \rightarrow \bar{n}}\equiv P_{n \rightarrow \eta} P_{\eta \rightarrow \eta} P_{\eta \rightarrow \bar{\eta}} P_{\bar{\eta} \rightarrow \bar{\eta}} P_{\bar{\eta} \rightarrow \bar{n}}$, where $P_{\eta \rightarrow \eta}$ and $P_{\bar{\eta} \rightarrow \bar{\eta}}$ are the survival probability of the $\eta$ and $\bar{\eta}$ particles, respectively. According to the CPT symmetry, $P_{\eta \rightarrow \eta}$ and $P_{\bar{\eta} \rightarrow \bar{\eta}}$ should be equal and satisfy the exponential decay law: $P_{\eta \rightarrow \eta} \equiv P_{\bar{\eta} \rightarrow \bar{\eta}} \equiv \exp(-\Gamma_{\eta}t)$, where $\Gamma_{\eta} \equiv \Gamma_{2}$ is the decay rate of $\eta$ and it is associated with its lifetime by $\tau_{\eta} \equiv 1/\Gamma_{\eta}$. If we assume $P_{\eta \rightarrow \eta}  \equiv P_{\bar{\eta} \rightarrow \bar{\eta}} \lesssim 10^{-7}$, the inconsistency can be explained. This requires that the lifetime of $\eta$ satisfies the condition:
\begin{equation}
\tau_{\eta} \equiv \frac{1}{\lambda_{\eta}} \lesssim - \frac{\tau_m}{\ln{\Big(\frac{P_{n\rightarrow \bar{n}}\Gamma_n^2}{\Gamma_a^2}\Big)}},
\end{equation}
where $\tau_m \simeq 0.1$ s is the mean propagation time of neutron in the ILL experiment \cite{baldo1994new}.  
If the lifetime of $\eta$ satisfies $\tau_{\eta} \lesssim 2.0 \times 10^{-3}$ s, the interpretation for the measurement of the neutron lifetime and the interpretation for the measurement of the $n$-$\bar{n}$ oscillation time with regard to the $n$-$\eta$ mixing can be consistent and the neutron lifetime anomaly can be explained in a direct and simple way. Hence, we could have a unified interpretation of the neutron lifetime anomaly and the $n$-$\bar{n}$ oscillation measurements based on the $n$-$\eta$ mixing. Note the above statements are given with regard to free $\eta$ particles. Similar to the reason for the stability of the neutron inside nuclei, the stability of the $\eta$ particle bounded in nuclei can be guaranteed by imposing additional assumptions or symmetries.

\subsection{Geometric phase}

Geometric phases, which provide a powerful tool for a unified description of the classical and quantum phenomena \cite{anandan1992geometric}, can be observed in a number of ways, such as polarized neutron interference (see e.g. Ref. \cite{hasegawa2001off}), vibrational spectroscopy (see e.g. Ref. \cite{von1998unambiguous}), and etc. In this work, we, specifically, consider the geometric phase associated with particle oscillations

The geometric phase and its observability has been discussed in the neutrino oscillation case \cite{capolupo2018geometric,lu2021comment,johns2022geometric,capolupo2022on}, where controversy has  emerged concerning the measurability of the Majorana phase and its connection to the geometric phase. The authors of Refs. \cite{capolupo2018geometric,capolupo2022on} argued that the Majorana phase can non-trivially contribute to a special type of the geometric phase defined in Refs. \cite{mukunda1993quantum,mukunda1993quantum2} and such a geometric phase may be measurable in neutrino oscillations. On the contrary, the author of Ref. \cite{lu2021comment} argued that the corresponding results presented in Ref. \cite{capolupo2018geometric} are not gauge-invariant and the Majorana phase can be eliminated from the geometric phase through a non-physical field rephrasing transformation (see also Ref. \cite{giunti2010no}), making it unlikely to be observed in neutrino oscillations. Shortly afterwards, the author of Ref. \cite{lu2021comment} commented on the assertions made in Refs. \cite{capolupo2018geometric,lu2021comment} and analyzed the gauge-invariant property of the off-diagonal geometric phase \cite{manini2000off} in neutrino oscillations. Recently, the authors of Refs. \cite{capolupo2018geometric,capolupo2022on} have replied to the comments given by the authors of Refs. \cite{lu2021comment,johns2022geometric} and explained why their arguments are reasonable. Since neutrinos have a tiny mass and only interact with matter very weakly, they are notoriously difficult to detect in experiments. This imposes a great challenge for the detection of the geometric phase in the neutrino sector and hence no evidence for such a geometric phase  has been found in neutrino oscillations so far.

In Sec. \ref{sec4}, we have discussed the observability of the Majorana phase associated with the $n$-$\bar{n}$ oscillations. Since the observability of the geometric phase is not necessarily determined by its dependence on the Majorana phase, in this work we only focus on the observable consequences of the geometric phase associated with the $n$-$\eta$ mixing, rather than attempting to resolve the controversy on the Majorana phase. Comparing with neutrinos, neutrons have a much larger mass and interact more strongly with matter, making it more feasible to detect the geometric phase in the neutron sector. The measurements of the geometric phase with neutrons have been suggested and conducted over the past decades \cite{richardson1988demonstration,weinfurter1990measurement,wagh1997experimental,wagh1998neutron,wagh2000neutron,hasegawa2001off}. Since the mutual transitions between $n$ and $\eta$ can be resulted from the $n$-$\eta$ mixing, a path-dependent geometric phase can be induced when neutrons propagate through space.

The geometric phase can be possibly observed through the following neutron interference experiment. A beam of highly coherent neutrons from a neutron source can be split into two neutron beams by a beam splitter. The two neutron beams travel inside the vacuum cavities of the two arms. The two arms have different lengths, i.e. $L_{1}$ and $L_{2}$, respectively. When the two neutron beams arrive at the same point of the detector, they can be recombined to produce interference. If the geometric phase is non-zero, any difference between the two arm lengths can give rise to interference effects between the two neutron beams. The measurement of the geometric phase induced by the $n$-$\eta$ mixing may provide another opportunity for the study of new physical effects.

\section{Conclusion}

In this work, we have explored the possibility that neutron ($n$) mixes with elementary neutral particle ($\eta$), which may have a non-zero lepton number ($\mathcal{L}=1$) and its decay products can be dark matter candidates. The $n$-$\eta$ mixing violates both the $\mathcal{B}$ and $\mathcal{L}$ symmetries by one unit, but conserves their difference $(\mathcal{B}-\mathcal{L})$. Furthermore, it is also featured by the mixing between composite and elementary particles and may give rise to non-trivial observable effects associated with the $n$-$\bar{n}$ oscillation that are different from the Standard Model predictions. We focus on two different scenarios, i.e. the  type-\RomanNumeralCaps{1} and type-\RomanNumeralCaps{2} oscillations, roughly corresponding to whether the $n$-$\eta$ mixing contributes to the absorptive mixing amplitude and whether the interference between oscillation and decay occurs. We have shown that such a mixing can serve as a versatile platform where many interesting phenomena occur and the investigations on such phenomena may open a promising avenue for exploring new physics beyond the SM.

In the scenario where the neutral particle oscillation is type-\RomanNumeralCaps{1}, the Majorana phase, which leads to CP-violating effects, can be observable. This is different from the situation in the type-\RomanNumeralCaps{2} oscillation, where the Majorana phases of the dispersive and absorptive parts cancel out and there would be no CP-violation unless some conditions are satisfied \cite{mckeen2016c}. $\mathcal{B}$-violation and CP-violation (along with C-symmetry violation) are two of the three conditions presented by Sakharov to explain the observed matter-antimatter asymmetry in our Universe \cite{sakharov1967violation}. The $n$-$\bar{n}$ oscillation induced by the $n$-$\eta$ mixing can be featured by both $\mathcal{B}$-violation and CP-violation and thus may open a promising window for future studies of matter-antimatter asymmetry. Moreover, in this scenario, the lower limits imposed by the results of the searches for $n$-$\bar{n}$ oscillations on the mass of the color multiplet boson (i.e. the new physics energy scale) have been presented. The derived constraints on the mass of the color multiplet boson varies gently respectively from $5$ to $25$ TeV and from $1$ to $8$ TeV throughout the entire range of the allowed $m_{\eta}$ values except in the vicinity of the neutron mass $m_n$. The derived new physics energy scales can be accessible to a direct detection at the LHC or future high-energy experiments \cite{allahverdi2013natural,deppisch2015neutrinos}. If the $n$-$\bar{n}$ oscillation was observed, the corresponding new physics particles, namely the color multiplet bosons, would be within the reach of direct searches at the LHC or future high-energy experiments. In this regard, the searches for the $n$-$\bar{n}$ oscillations can provide a complementary and economical way of searching for new physics besides the direct searches for new physics at high-energy colliders.

In the scenario where the neutral particle oscillation is type-\RomanNumeralCaps{2}, we analyze the testable implications on masses and lifetimes. The $n$-$\bar{n}$ oscillation induced by the $n$-$\eta$ mixing gives rise to two mass eigenstates, which are predicted to have different masses and lifetimes. One mass eigenstate ($N_1$) is predominantly composed of neutron state ($n$) and the other one ($N_2$) is predominantly composed of anti-neuron state ($\bar{n}$). The constraint imposed by the experimental searches for the $n$-$\bar{n}$ oscillations on the lifetime difference is predicted to be as large as $1.84 \times 10^{-3}$ s, which may be within the detectable regions of future experiments. In the SM, the commonly recognized neutrons, which almost inclusively decay into electron, proton, and antineutrino through the $\beta$-decay process ($n \rightarrow pe^{-}\bar{\nu}_e$) \cite{zyla2020review}, might not necessarily be mass eigenstates in the presence of exotic interactions and thus might not be described properly in the conventional treatment of the SM. For example, in the presence of the $n$-$\eta$ mixing, a commonly recognized neutron, which is created through the weak and strong interactions, might not be a mass eigenstate and thus might not have a well-defined lifetime. In this case, we could explore the compatibility between the interpretation of the neutron lifetime anomaly and the interpretation of the $n$-$\bar{n}$ oscillation experiments in connection with the $n$-$\eta$ mixing. If the lifetime of $\eta$ satisfies the condition: $\tau_{\eta} \lesssim 2.0 \times 10^{-3}$ s, a unified interpretation of the two types of experiments based on the same $n$-$\eta$ mixing can be suggested. In this manner, the neutron lifetime anomaly can be explained in a direct and simple way.

In both scenarios, the bounds on the mass of the color multiplet bosons depend on the size of the coupling constants, whereas some bounds on the coupling constants imposed by the FCNC measurements can be more restrictive than the typical values we choose. In this case, a balance between all the experimental constraints needs to be considered. We could adjust the mass of $\eta$ so that our results are consistent with the FCNC constraints as well as with the results of the direct searches for the color multiplet bosons at the LHC. In order to satisfy all the constraints, the mass of $\eta$ should be close to the mass of the neutron and thus cannot be randomly chosen.

Finally, we have discussed about the observability of the geometric phase associated with the $n$-$\eta$ mixing. The measurement of such a geometric phase may provide another opportunity for the study of the new physical effects. Comparing with neutrinos, neutrons have a much larger mass and interact more strongly with matter, making it more feasible to detect such a geometric phase through a neutron interference experiment and a possible measurement scheme has also been suggested.

\begin{appendices}

\section{\label{appa}}

The effective mass matrix $W$ can be diagonalized by the transformation matrix $T$:
\begin{equation}
\begin{split}
T W T^{-1} =& 
\left[
\begin{array}{cc}
c_1 & \epsilon_1 \\
\epsilon_2 & c_2
\end{array}
\right]
\left[
\begin{array}{cc}
W_{11} & W_{12}\\
W_{21} & W_{22}
\end{array}
\right]
\left[
\begin{array}{cc}
c_1 & \epsilon_1 \\
\epsilon_2 & c_2
\end{array}
\right]^{-1}\\
=&
\left[
\begin{array}{cc}
\omega_{1} & 0\\
0 & \omega_{2}
\end{array}
\right].    
\end{split}
\end{equation}
Here, the mass and width of the mass eigenstates $\ket{n_1}$ and $\ket{n_2}$ are given by
\begin{align}
&\begin{aligned}
\omega_{1,2}= \frac{1}{2} \left[M_{11}+M_{22} \pm \Re(\Delta W) \right],
\label{eq6}
\end{aligned}\\
&\begin{aligned}
\Gamma_{1,2} =\frac{1}{2} \left[ \Gamma_{11} + \Gamma_{22} \mp 2\Im\big(\Delta  W \big)\right],
\label{eq7}
\end{aligned}
\end{align}
with
\begin{equation}
\Delta  W \equiv \left[\left(W_{11}-W_{22}\right)^2 - 4 W_{12} W_{21}\right]^{\frac{1}{2}}.
\label{deltaW}
\end{equation}
In this work, unless otherwise specified, we assume that the CPT symmetry is conserved and thus particle and anti-particle have the same mass, i.e. $m_{n}=m_{\bar{n}}$ and $m_{\eta}=m_{\bar{\eta}}$. If the condition: $|W_{11}| \gtrsim  |W_{22}| \gg |W_{12}| \simeq  |W_{21}|$ is satisfied, the mixing coefficients will satisfy the following conditions:
\begin{align}
&\begin{aligned}
c \equiv c_1 \simeq  c_2, \quad |c|  \simeq 1,
\label{c11}
\end{aligned}\\
&\begin{aligned}
\epsilon \equiv \epsilon_1 \simeq  -\epsilon_2, \quad |\epsilon|  \ll 1.
\label{e11}
\end{aligned}
\end{align}
With the above approximations, the transformation matrix $T$ can be written as
\begin{equation}
T
=
\left[
\begin{array}{ccl}
\cos\theta &\sin\theta\\
-\sin\theta &\cos\theta
\end{array}
\right].
\end{equation}
Here, $\theta$ is the mixing angle.

Note this is not the unique approach to diagonalize the effective mass matrix $W$. The bi-unitary transformation can also be used to diagonalize the effective mass matrix \cite{haber2021tale, choi2011mathematics}. In the bi-unitary transformation, the obtained diagonal elements are not necessarily eigenvalues but instead they are singular values (for more details, please see Refs. \cite{haber2021tale, choi2011mathematics}). Since we intend to obtain the eigenvalues of the effective mass matrix, we choose the conventional transformation described above throughout this work.

\begin{table*}[t]
\caption{Comparison of the pure and impure oscillation within the formalism of the SM.
}
\begin{ruledtabular}
\begin{tabular}{l|c|c}
\diagbox{Property}{Type} & (\RomanNumeralCaps{1}) Pure oscillation & (\RomanNumeralCaps{2}) Impure oscillation\\\hline
Examples& neutrino oscillation ($\nu_e \rightleftarrows \nu_\mu \rightleftarrows \nu_\tau$) & $K^0$-$\bar{K}^0$, $B^0$-$\bar{B}^0$, $D^0$-$\bar{D}^0$ oscillation\\\hline
Interference effects& no interference between oscillation and decay & has an interference between  oscillation and decay\\\hline 
Mixing angles&are free parameters and cannot be calculated & can be calculated via loop-level diagrams\\\hline
Origin& unknown & mainly from the weak interactions\\
\end{tabular}
\end{ruledtabular}
\label{appctab}
\end{table*}

\section{\label{appb}}

In the presence of external magnetic fields, the probabilities for the $n$-$\bar{n}$ and $\bar{n}$-$n$ oscillations without making the approximation associated with the limit $\phi_{1,2} \gg 1$ can be given by Eq. (\ref{fullpnn}) and (\ref{fullpnn2}) respectively. Note the direction of the magnetic field in Eq. (\ref{fullpnn2}) is opposite to the direction of the magnetic field in Eq. (\ref{fullpnn}). Eq. (\ref{fullpnn}) and (\ref{fullpnn2}) show explicitly that the Majorana phase can not only be observable but also give rise to a CP-violating effect.

\begin{widetext}
\begin{equation}
\begin{split}
P_{n \rightarrow \bar{n}}^{B \neq 0} \simeq & \sin^2(2\theta_1)\sin^2\Big(\frac{\phi_1}{2}\Big) \Big[\Big(\frac{m_\eta}{m_n}\Big)^2 \cos^4\theta_1 + \sin^4\theta_1 + \frac{1}{2} \Big( \frac{m_\eta}{m_n} \Big)  \sin^2(2\theta_1) \cos(\phi_1 + 2\xi )\Big] \sin^2(2\theta_2)\sin^2\Big(\frac{\phi_2}{2}\Big)\\
= & \frac{1}{2} \Big\{ \frac{4 \delta^2  \sin^2\Big[\frac{\sqrt{(m_n - \vert \mu_{n} B \vert -m_{\eta})^2 + 4 \delta^2 }}{2} t \Big]}{(m_n-|\mu_n B|-m_{\eta})^2+ 4\delta^2} \Big\} \Big\{\Big( \frac{m_{\eta}}{m_n} \Big)^2 \Big[\frac{(m_n-|\mu_n B|-m_{\eta})^2 +2 \delta^2}{(m_n-|\mu_n B|-m_{\eta})^2+ 4\delta^2} + \sqrt{ \frac{(m_n-|\mu_n B|-m_{\eta})^2}{(m_n-|\mu_n B|-m_{\eta})^2+ 4\delta^2} } \Big] \\
&+\frac{1}{2}\Big( \frac{m_{\eta}}{m_n} \Big) \frac{4 \delta^2 \cos(\sqrt{(m_n - \vert \mu_{n} B \vert -m_{\eta})^2 + 4 \delta^2 } t + 2\xi)}{(m_n-|\mu_n B|-m_{\eta})^2+ 4\delta^2} + \Big[\frac{(m_n-|\mu_n B|-m_{\eta})^2+2 \delta^2}{(m_n-|\mu_n B|-m_{\eta})^2+ 4\delta^2} \\
&- \sqrt{ \frac{(m_n-|\mu_n B|-m_{\eta})^2}{(m_n-|\mu_n B|-m_{\eta})^2+ 4\delta^2} }\Big] \Big\} \Big\{ \frac{4 \delta^2 \sin^2 \Big[\frac{\sqrt{(m_n + \vert \mu_{n} B \vert -m_{\eta})^2 + 4 \delta^2 }}{2} t \Big]}{(m_n+|\mu_n B|-m_{\eta})^2+ 4\delta^2} \Big\}.    
\end{split}
\label{fullpnn}
\end{equation}

\begin{equation}
\begin{split}
P_{\bar{n} \rightarrow n}^{B \neq 0} \simeq & \sin^2(2\theta_1)\sin^2\Big(\frac{\phi_1}{2}\Big) \Big[\Big(\frac{m_\eta}{m_n}\Big)^2 \cos^4\theta_1 + \sin^4\theta_1 + \frac{1}{2} \Big( \frac{m_\eta}{m_n} \Big)  \sin^2(2\theta_1) \cos(\phi_1 - 2\xi )\Big] \sin^2(2\theta_2)\sin^2\Big(\frac{\phi_2}{2}\Big)\\
= & \frac{1}{2} \Big\{ \frac{4 \delta^2  \sin^2\Big[\frac{\sqrt{(m_n - \vert \mu_{n} B \vert -m_{\eta})^2 + 4 \delta^2 }}{2} t \Big]}{(m_n-|\mu_n B|-m_{\eta})^2+ 4\delta^2} \Big\} \Big\{\Big( \frac{m_{\eta}}{m_n} \Big)^2 \Big[\frac{(m_n-|\mu_n B|-m_{\eta})^2 +2 \delta^2}{(m_n-|\mu_n B|-m_{\eta})^2+ 4\delta^2} + \sqrt{ \frac{(m_n-|\mu_n B|-m_{\eta})^2}{(m_n-|\mu_n B|-m_{\eta})^2+ 4\delta^2} } \Big] \\
&+\frac{1}{2}\Big( \frac{m_{\eta}}{m_n} \Big) \frac{4 \delta^2 \cos(\sqrt{(m_n - \vert \mu_{n} B \vert -m_{\eta})^2 + 4 \delta^2 } t - 2\xi)}{(m_n-|\mu_n B|-m_{\eta})^2+ 4\delta^2} + \Big[\frac{(m_n-|\mu_n B|-m_{\eta})^2+2 \delta^2}{(m_n-|\mu_n B|-m_{\eta})^2+ 4\delta^2} \\
&- \sqrt{ \frac{(m_n-|\mu_n B|-m_{\eta})^2}{(m_n-|\mu_n B|-m_{\eta})^2+ 4\delta^2} }\Big] \Big\} \Big\{ \frac{4 \delta^2 \sin^2 \Big[\frac{\sqrt{(m_n + \vert \mu_{n} B \vert -m_{\eta})^2 + 4 \delta^2 }}{2} t \Big]}{(m_n+|\mu_n B|-m_{\eta})^2+ 4\delta^2} \Big\}.    
\end{split}
\label{fullpnn2}
\end{equation}
\end{widetext}

\section{\label{appac}}

Tab. \ref{appctab} summarizes the main differences between the pure and impure oscillation.

\end{appendices}

%\clearpage

\section*{Acknowledgement}
This work is supported by the National Natural Science Foundation of China (Grant No. 12022517), the Science and Technology Development Fund, Macau SAR (File No. 0048/2020/A1). The work of Yongliang Hao is supported by the National Natural Science Foundation of China (Grant No. 12104187), Macao Youth Scholars Program (No. AM2021001), Jiangsu Provincial Double-Innovation Doctor Program (Grant No. JSSCBS20210940), and the Startup Funding of Jiangsu University (No. 4111710002).

\clearpage

\bibliography{nnbar}
\end{document}